\RequirePackage{fix-cm}
\documentclass [smallcondensed,final] {svjour3}
\smartqed  
\usepackage{graphicx}
\usepackage{multirow}
\usepackage{xcolor}
\usepackage[T1]{fontenc}
\usepackage[utf8]{inputenc}
\usepackage[hyphens]{url}
\usepackage{amsmath}
\usepackage{cite}
\usepackage[hidelinks]{hyperref} 
\journalname{Journal of Infrared, Millimeter, and Terahertz Waves}
\begin{document}

\title{THz detection of biomolecules in aqueous environments – status and perspectives for analysis under physiological conditions and clinical use
}

\titlerunning{THz detection of biomolecules in aqueous environments}        

\author{Christian~Weisenstein \and Anna~Katharina~Wigger \and Merle~Richter \and Robert~Sczech \and Anja~Katrin~Bosserhoff \and Peter~Haring~Bol\'ivar}

\institute{C. Weisenstein, A. K. Wigger, M. Richter, R. Sczech and P. Haring Bol\'ivar \at
              Institute of High Frequency and Quantum Electronics, \\
              University of Siegen, H\"olderlinstrasse 3, 57076, Siegen, Germany.\\
              Tel.: +49 271 740-2904\\
              Fax:  +49 271 740-2410\\
              \email{christian.weisenstein@uni-siegen.de, anna.wigger@uni-siegen.de,\\ 
              merle.richter@uni-siegen.de, robert.sczech@uni-siegen.de,\\ 
              peter.haring@uni-siegen.de}           
           \and
           A. Katrin Bosserhoff \at
              Institute of Biochemistry and Molecular Medicine,\\
              Friedrich Alexander University Erlangen-N\"urnberg,\\ 
              Fahrstrasse 17, 91054, Erlangen, Germany.\\
              \email{anja.bosserhoff@fau.de}
}

\date{Received: 08.01.2021 / Accepted: 07.04.2021}

\maketitle
\begin{abstract}
Bioanalytical THz sensing techniques have proven to be an interesting and viable tool for the label-free detection and analysis of biomolecules. However, a major challenge for THz bioanalytics is to perform investigations in the native aqueous environments of the analytes. This review recapitulates the status and future requirements for establishing THz biosensing as a complementary toolbox in the repertoire of standard bioanalytic methods. The potential use in medical research and clinical diagnosis is discussed. Under these considerations, this article presents a comprehensive categorization of biochemically relevant analytes that have been investigated by THz sensing techniques in aqueous media. The detectable concentration levels of ions, carbohydrates, (poly-)nucleotides, active agents, proteins and different biomacromolecules from THz experiments are compared to characteristic physiological concentrations and lower detection limits of state-of-the-art bioanalytical methods. Finally, recent experimental developments and achievements are discussed, which potentially pave the way for THz analysis of biomolecules under clinically relevant conditions.
\end{abstract}

\section{Introduction}
\label{intro}
Over the past three decades, significant advances in the development of widely usable THz systems have generated a plethora of  research activities by a rapidly growing THz community in numerous application fields. For several years, the THz community has demonstrated in a large number of articles, that in the electromagnetic spectrum, the long-wavelength end of the far-infrared range shows a huge potential for industrial utilisation and scientific research \cite{Mittleman1996, Smith2011, Jepsen2011, Amenabar2013}. The combination of low-energy radiation with the terahertz transparency of various materials provides ample opportunities for versatile fields of applications \cite{Ferguson2002, Tonouchi2007, Fischer2013}. Due to the characteristic resonances of many biomolecules at THz frequencies, life sciences are a major target field of THz research. In this respect, one of the primary objectives is the development of analytical tools for biomolecular research and disease diagnostics. In addition to the potential application in the bioanalytical field \cite{Zhang2002, Siegel2002, Taylor2011}, in life science\textcolor{black}{s} there is a general cross-platform interest to quantitatively record and qualitatively describe physiological relevant analytes, in particular functional units like proteins. Therefore, the main aim is the detection of a desired molecule and the characterization of its properties and interactions in the native and complex biomolecular network \cite{Hansma1997, Patton2002, Fan2008}. In this overview, we concentrate on an assessment and evaluation of the current state of THz-based bioanalytic experimental tools, which show a potential application in the fields of clinical diagnostics and medical research.

\textcolor{black}{The main part of the article presents a review of the minimal detected concentration of physiologically relevant analytes in their native aqueous environments that have been investigated by state-of-the-art THz measurement techniques. Although the determination of the lower detection limit (LDL) in the sense of its precise definition requires more information on the chemical measurement process for each technique and analyte involving for instance elaborate serial dilution experiments \cite{IUPACGoldBook, IUPACOrangeBook, Currie1995}, the minimal reported concentration will be referred to as attained LDL over the course of this article, as the wide majority of THz analysis to date do not perform dilution experiments.} Presenting a systematic categorization of biomarkers and considering representative analytes, the \mbox{(patho-)}physiological relevance of these biomarkers \textcolor{black}{is} defined and typical physiological concentrations or typical reference ranges are provided. Subsequently, the sensitivities measured by THz techniques are compared to established state-of-the-art bioanalytical methods. This comparison helps to set a quantitative target to increase both the sensitivity and the \textcolor{black}{selectivity} of THz sensing techniques. Evidently, this is essential in order to establish them as a competitive label-free alternative detection technique, which can be appropriate for \textit{in vivo} and real-time application in clinical diagnostics and medical research.

In the last part of the article, we invite the THz community to participate in this critical discussion, assessing the current state-of-the-art of THz sensing of biomolecules in native aqueous environments. Valuable experimental insights reflect important steps and partially outstanding contributions towards THz analysis of biomolecules in aqueous environments under physiological conditions. Nevertheless, recommendations on the prospective THz sensing of biomolecules in absorbing/aqueous media are proposed. These recommendations are intended to promote the broader use of THz radiation in the biomedical and clinical context. 

\section{Experimental approaches for THz sensing of biomolecular analytes in aqueous environments} 
In this section, we review and evaluate the potential of THz sensing techniques for the detection and characterization of biomolecular compounds in their native absorbing environments. The following evaluation is formatted as a dialogue in order to illustrate the varied aspects and requirements which need to be addressed before a widespread application of THz techniques can be envisaged.\\

\textbf{Which analytes are of fundamental interest, independently from the sensing platform?} \\
Biomolecular analytes of fundamental interest have a significant (patho-)\-physio\-logical relevance and are  accessible to analytic procedures and instruments. Therefore, their occurrence in an easily accessible natural context like body fluids is of paramount importance and is a major concern in clinical diagnostics. Furthermore, it is interesting to investigate biomolecules that are associated with specific biochemical functionalities (i.e. \emph{biomarkers}). Molecules like enzymes, antibodies, other proteins, DNA or biomacromolecules in general are often characterized by a higher degree of structural complexity closely associated with their unique and \textcolor{black}{selective} biochemical function. \\

\textbf{Which established techniques do support the quantitative detection and qualitative description of biomacromolecules?} \\
There is a wide variety of established bioanalytical tools that can be used for molecular detection and characterization. \textcolor{black}{The application of these techniques in a general sense are summarized in table\,\ref{tabTechniques}}.

\begin{table}[h]
\begin{center}
\centering
\caption{\textcolor{black}{Selection of the main established techniques for the quantitative detection and qualitative description of biomacromolecules.}}
\begin{tabular}{ |l|l| } 
\hline
\multirow{6}{7em}{Identification of molecules} & -- Matrix-assisted laser desorption/ionization-time-of-flight \\
& (MALDI-TOF) \cite{Yates1998} \\
& -- High-performance liquid chromatography (HPLC) \cite{Yates1998}  \\ 
& -- UV-Vis spectroscopy \cite{brown2009}  \\ 
& -- Immunoassays \cite{Luppa2001}  \\ 
& -- Affinity chromatography \cite{Jin2002}  \\ 
\hline
\multirow{3}{7em}{Evaluation of size and mass} & -- Mass spectrometry \cite{Yates1998} \\ 
& -- Gel chromatography \cite{Maire1980}  \\ 
& -- Centrifugation \cite{Lebowitz2002}  \\ 
\hline
\multirow{3}{7em}{Observation of the shape and structure} & -- X-ray diffraction (XRD) analysis \cite{Uson1999} \\ 
& -- Nuclear magnetic resonance (NMR) \cite{Kainosho2006}  \\ 
& -- Circular dichroism (CD) \cite{Chen1972}  \\ 
\hline
\multirow{3}{7em}{Investigation of the function} & -- Ligand-binding assay via surface plasmon resonance (SPR) \cite{Homola2008} \\ 
& -- Comparing the amino acid sequence with entries in databases such\\
& as basic local alignment search tool (BLAST) \cite{Altschul1990} \\ 
\hline
\multirow{2}{7em}{Determination of the quantity} & -- Enzyme-linked immunosorbent assay (ELISA) \cite{Lequin2005} \\ 
& -- Two-dimensional polyacrylamide gel electrophoresis (2D-PAGE) \cite{Issaq2008}  \\ 
& \textcolor{black}{-- Western Blotting \cite{Yang2012}}\\
\hline
\end{tabular}
\label{tabTechniques}
\end{center}
\end{table}

While the described bioanalytical tools focus mainly on the detection and characterization of proteins, other \textcolor{black}{selective} analytical techniques are available for the analysis of nucleic acids (e.g., reverse transcription polymerase chain reaction (RT-PCR) for the qualitative detection of gene expression products \cite{Bustin2005}, DNA sequencing techniques \cite{Shendure2008} and fluorescence-based DNA-microchip arrays \cite{Shalon1996}). 
 
As shown by the immense literature on biosensing techniques \cite{turner1987biosensors, holzinger2014nanomaterials} there are many established and emerging bioanalytical methods that can be used for the detection and characterization of biomolecules. \\

\textbf{Which opportunities  can be exploited by THz radiation for detecting and characterizing biomolecules?} \\
In comparison to established bioanalytical techniques, important characteristics of THz technology are provided by intrinsic biomolecular resonances that can be excited and detected at THz frequencies \cite{powell1987investigation, van1989millimeter, walther2000far, brucherseifer2000label}. The capability of far infrared electromagnetic radiation to induce intra- and intermolecular oscillations is a unique advantage and a prominent attribute of macromolecules due to the excitation of vibrational modes that are delocalized over a large group of atoms \cite{brucherseifer2000label, BornPhD2010}. This fact is specifically interesting for molecular characterization under physiological conditions because the molecules permanently interact with their complex environment. One important aspect of these interactions are hydrogen bonds whose dynamics can be observed along with hydration dynamics \cite{brucherseifer2000label, BornPhD2010}. This may help in decoding complex molecular functionalities and reaction mechanisms. Based on this, THz techniques could uniquely and significantly contribute in developing new tools in biomedical diagnostics in order to gain information about unknown or complicated diseases such as cancer and neurodegenerative or cardiovascular disorders. 

Apart from a desired reliable quantitative determination and characterization of molecular binding events, there is an ongoing demand for developing therapeutic applications, too \cite{Titova2013, son2019potential}. Generally, THz radiation would be showing its complete potential when applied, for instance, in the label-free and real-time monitoring of therapeutically relevant analytes within complex matrices in vivo to ensure accurate dosage during chemotherapeutical treatments \cite{Kelland2007}. In a scenario closer to realization, THz radiation may be used for assisting probing the effectiveness of binding events of therapeutic agents of purified samples. Yet, some challenges remain until these visions can be achieved.\\

\textbf{What aspects aggravate analyses of biomolecules in aqueous media?} \\
Many interesting target molecules are embedded in intracellular processes which are generally difficult to access with any direct bioanalytical method. This may always demand sample preparation and purification to some extent prior to analysis. Apart from this, there are two main issues preventing widespread THz-based detection of biomolecules. First, the THz excitation of large biomolecular entities generates many excited vibrational modes simultaneously. As a result, both the non-specificity of the THz excitation and the THz response generally tend to increase the inhomogeneous broadening of the spectrum with increasing intra- and intermolecular sample complexity and\,/\,or disorder \cite{Laman2008}. \textcolor{black}{The  selectivity, or - in its ultimate form - specificity, of a technique describes the extent to which other stimuli (chemical or physical) can interfere with an analyte's signal. If such a signal cannot be explicitly assigned to one analyte species, the corresponding method must be denoted as "non-specific" or even "non-selective". As mentioned above, THz radiation has the capability to excite a multitude of e.g. vibrational or rotational modes which in many cases are not distinguishable and will therefore be referred to as the "non-specificity of THz excitation" within this manuscript \cite{IUPACGoldBook, DenBoef1983}.} This effect depends on the molecular conformation and crystallinity. Although there is a good possibility of observing the spectral features of small molecules \cite{Peiponen2012}, the spectral features of large molecules tend to vanish in the broadened THz spectral response \cite{Markelz2000a}. Consequently, it becomes difficult to distinguish complex molecules that exhibit similar structures. This directly affects the THz detection and characterization potential. In addition, some important biomolecular investigations are performed under physiological conditions that are significantly influenced by the molecular interplay \cite{Born2009a}. The term physiological indicates inter alia in aqueous environments leading to a second obstacle of investigation techniques in the THz range. Water in liquid form exhibits a high absorption coefficient at THz frequencies \cite{LIEBE1991, Kindt1996}. Therefore, the dominant dielectric response of water can easily mask the analyte´s dielectric response. To counteract this problem and in order to discriminate both the spectral responses of the biomolecular aqueous solution and water, highly concentrated sample solutions are presently needed. Consequently, as shown in more detail in Sect.~III, the values of typical sample concentrations (or densities, etc.) measured so far are well beyond physiological concentrations and typically close to saturation. Depending on the experiment design, the dielectric response can also be affected by overall charge \cite{Qiao2012} or partially be a function of the significant water displacement \cite{Laurette2012}. At this point, it must be emphasized that in contrast to the term "physiological concentration", to perform an experiment under physiological conditions requires the mimicry of an \textit{in vivo} environment in all relevant factors such as, but not limited to temperature, pressure, pH, dielectric potential and presence of all components of a cell-like matrix.

In summary, the direct detection (by differentiation of vibrational modes) of complex biomolecules using available state-of-the-art THz sensing approaches is currently a challenging task due to high water absorption as well as low THz excitation and detection \textcolor{black}{selectivity}. \\

\begin{figure*}
\centering
\includegraphics[width=0.80\textwidth]{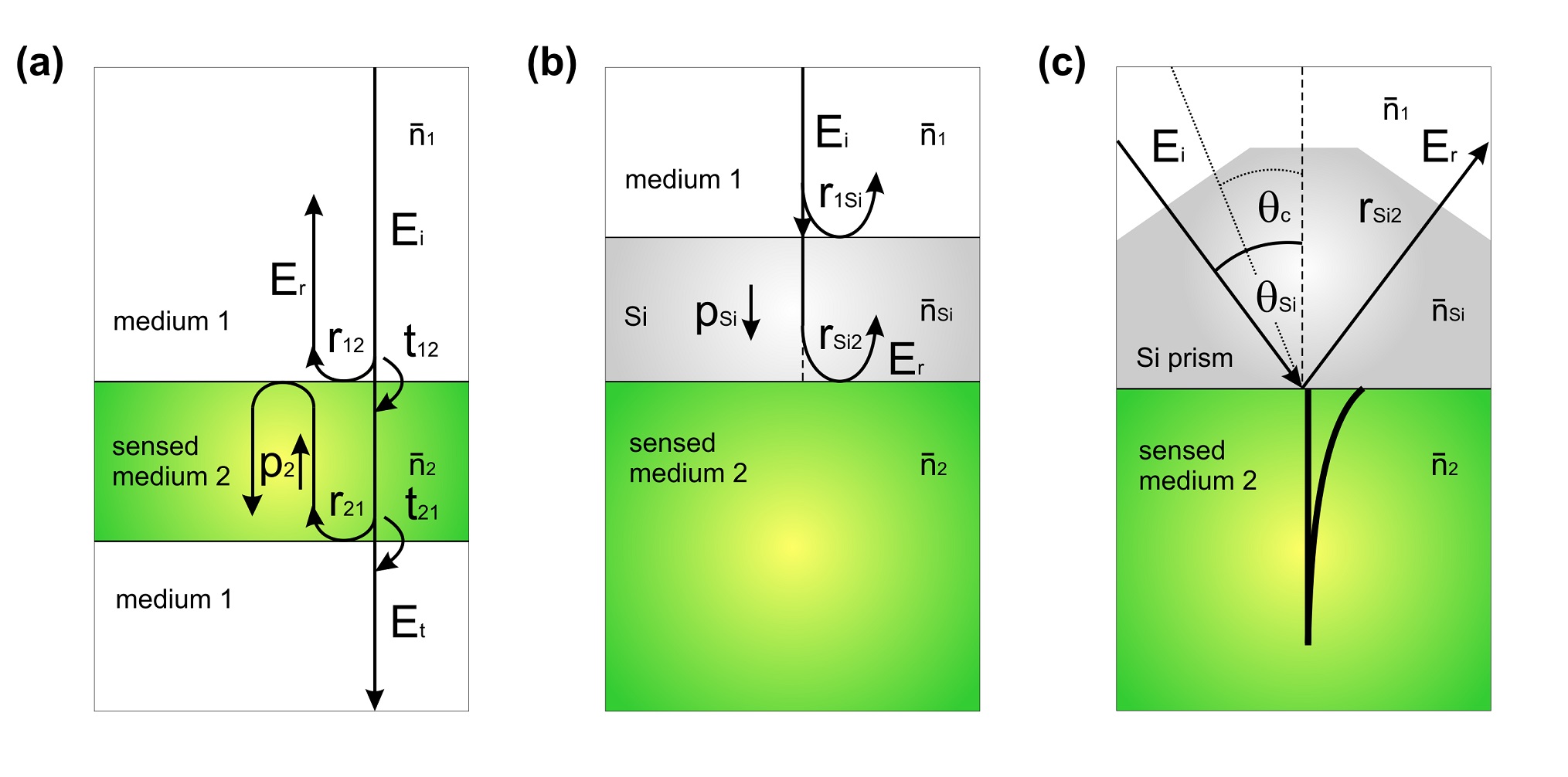}
\caption{Schematic of the traditionally employed THz sensing techniques for spectroscopy of absorbing liquids. (a) Transmission through a stratified film consisting of three homogeneous media with the complex refractive index $\bar{n}_1$, $\bar{n}_2$, and $\bar{n}_3$. The transmission (t) and reflection (r) coefficient indices represent the medium interfaces. Each completed field propagation through medium 2 is taken into account with the propagation term $p_2$. (b) The intensity of the reflected beam is changed according to the refractive index inside sensed medium 2. A self-referencing system is established by using additional interfaces introduced by a window (e.g., silicon). Expressions $r_{\text{Si2}}$ and $r_{\text{1Si}}$ are used to calculate the reflectivity $R$. (c) Attenuated total reflection (ATR): The incident field is totally reflected internally at the prism\,/\,sensed medium interface for angles $\theta_{\text{Si}}$ larger than the critical angle $\theta_c$. The evanescent field reduces the intensity of the reflected beam by penetrating and interacting with the solution. The reflectivity can be calculated by ATR\,=\,$r_{\text{Si2}}$ and $r_{\text{Siair}}$, which represents the reference situation.}
\label{TTF}
\end{figure*}

\textbf{Which THz techniques are traditionally applied for biomolecular detection and characterization in aqueous media?} \\
For THz investigation of polar liquids, such as aqueous solutions, or more specifically bioanalytes in their native aqueous environment, the transmission \cite{Kindt1996} and reflection \cite{Jepsen2007} setups as well as the modified schematic derivatives (i.e., attenuated total reflection \cite{Nagai2006}, transmission through reverse micelles that are distributed in non-polar solvents \cite{Mittleman1997} and waveguide-based sensors \cite{Ohkubo2006}) are the widely adopted techniques (Figure\,\ref{TTF}). \textcolor{black}{Other alternative techniques are being developed, such as THz ellipsometry approaches for the determination of refractive indices of liquids \cite{Otani2010,Azarov2018,Neethling2019}, which, however, have not yet been widely used for the detection of biomolecules. Furthermore, first experiments have been performed recently using a THz near-field setup for the detection of biomolecules \cite{Heo2020}.} For more technical details on the traditional THz sensing approaches of transmission and reflection we would like to refer to a comprehensive review article from Jepsen et al. \cite{Jepsen2011}. The following paragraphs summarize the analytical formulas that describe such experiments in order to allow quantitative sensitivity assessment of the corresponding techniques.

Throughout the article, we refer to the Fresnel transmission ($t_{mn}^{p}$) and reflection ($r_{mn}^{p}$) coefficients for a p-polarized plane-wave incident at an interface that separates two homogeneous media $m$ and $n$ determined by

\begin{flalign}
\> r_{mn}^{p} &= \frac{E_r}{E_i} = \frac{\bar{n}_mcos(\theta_n) - \bar{n}_ncos(\theta_m)}{\bar{n}_mcos(\theta_n) + \bar{n}_ncos(\theta_m)}, \label{Fresnel_r}\\
\> t_{mn}^{p} &= \frac{E_t}{E_i} = \frac{2\bar{n}_{n}cos(\theta_n)}{\bar{n}_ncos(\theta_m) + \bar{n}_mcos(\theta_n)}, \label{Fresnel_t} 
\end{flalign}

from which the ratio of the reflected (transmitted) and incident electric fields $E_r$ ($E_t$) and $E_i$ can be calculated, respectively. The complex refractive index is denoted by $\bar{n}$ while $\theta$ describes the angle between the incident and transmitted waves \cite{BornWolf2003}.

The transmission configuration is the most popular spectroscopic approach. In most cases, a thin cuvette is loaded with the liquid being analyzed that produces a homogeneous stratified medium as shown in Figure\,\ref{TTF}(a). Considering the Fresnel coefficients, the transmitted electric field for a sample ($E_{\text{sam}}$) and a reference using an empty cuvette ($E_{\text{ref}}$) can be expressed as

\begin{flalign}
\> E_{\text{sam}}(\omega) &= t_{12}p_2t_{23} \sum\limits^\infty_{k=0}[r_{12}p_2^2r_{23}]^kE_i(\omega), \label{eq:sam}\\
\> E_{\text{ref}}(\omega) &= t_{1a}p_at_{a3} \sum\limits^\infty_{k=0}[r_{1a}p_a^2r_{a3}]^kE_i(\omega). \label{eq:ref}
\end{flalign}

The sum term represents Fabry-P\'erot resonances and the propagation term is denoted by  

\begin{align}
p_2 = \text{exp}\{-i\bar{n}_2\omega d/c\},
\label{eq:prop_term}
\end{align}

where $\omega$ is the angular frequency, $d$ is the film thickness, and $c$ is the speed of light. 

Depending on the experimental conditions, Fabry{--}P\'erot resonances may occur inside the thin layers. The order of $k$ in Equations\,(\ref{eq:sam}) and (\ref{eq:ref}) refers to the number of the total forward and backward reflections inside the thin film. The impact of such etalon effects gains importance with decreasing layer thickness \cite{Duvillaret1996, duv99}. The complex field transmission coefficient of the sample $T(\omega)$ can be obtained from $E_{\text{sam}}(\omega)$/$E_{\text{ref}}(\omega)$ when both the amplitude and phase of the electromagnetic wave are recorded. In order to obtain the complex refractive index $\bar{n}_2 = n\,-\,i\kappa$, fitting algorithms can be used to numerically approach the measured value of $T(\omega)$ \cite{Ebeling1989}. 

The ideal thickness of the sample strongly depends on the absorption coefficient of the liquid under test considering the transmission configuration in Figure\,\ref{TTF}(a). The sampling of non-polar liquids helps extend the thicknesses of the sample in the range of millimeter to centimeter \cite{Pedersen1992, Keiding1997}. The thickness is usually reduced to d$\,\leq\,$200\,$\mu$m in the case of water and other high absorbing liquids \cite{Kindt1996, Venables2000, Asaki2002, Kitahara2005, Bergner2005, Saha2012, Yomogida2013}. A sophisticated approach to extend the propagation path length is shown by inverse micelles, which reduce the overall absorption of the system \cite{Mittleman1997, Boyd2002}. In order to characterize absorbing liquids, the general requirement of short propagation distances has also led to the development of wave guiding structures for tuning electromagnetic field liquid interactions \cite{Ohkubo2006, Cheng2008, sczech2012}.

An alternative to measure liquids that are difficult to access in a transmission configuration, is the THz reflection technique. As depicted in Figure\,\ref{TTF}(b), reflection helps establish a self-referencing system. One single scan can simultaneously detect two reflected pulses in the sample when a second interface is introduced (e.g., a silicon window). The first pulse can be used as a reference if both signals are separated to a sufficient degree from each other in time domain. Then, the reflectivity at normal incidence is calculated from $R\,=\,$$t_{\text{1Si}}$$t_{\text{Si1}}$$r_{\text{Si2}}$$p_{\text{Si}}^2$/$r_{\text{1Si}}$ with $\theta_n\,=\,0^{\circ}$. Using this approach (Figure\,\ref{TTF}\,(b)), one single THz scan can also help determine the optical constants of an analyte as introduced by \cite{Thrane1995, Ronne1997, Jepsen2008, Moller2009}.

Another alternative, terahertz attenuated total reflection, has been developed to analyze high absorbing materials \cite{Hirori2004}. A typical setup is shown in Figure\,\ref{TTF}\,(c). Contrary to the optical frequency range, as the penetration length roughly increases with the wavelength, the penetration depth of the evanescent field in the sensed medium becomes significantly larger in the THz regime. \textcolor{black}{The three system configurations presented are used, among other things, depending on the absorption coefficient and thickness of the sample. The measurement time can vary greatly depending on the type of measurement system used. For typical commercial available systems, it can very from 33\,ms per scan \cite{Kaun2005} to 36\,s per scan for a homemade THz-TDS system \cite{Takebe2013} independent from the measurement configuration. Faster measurement systems are also available, but it should be noted here that the absolute measurement time is not the decisive factor, since typical preparatory steps of biochemical analyses such as purification, immobilization or absorption take significantly more time.}

According to Figure\,\ref{TTF}\,(c), by applying Snell's law, Equation\,(\ref{Fresnel_r}) can be represented by the expression

\begin{flalign}
\> r_{\text{Si,2}} &= \frac{n_{\text{Si}} \sqrt{n_2^2 - n_{\text{Si}}^2 \text{sin}^2\theta_{\text{Si}}} - n_2\,\text{cos}\,\theta_{\text{Si}}} {n_{\text{Si}} \sqrt{n_2^2 - n_{\text{Si}}^2 \text{sin}^2\theta_{\text{Si}}} + n_2\,\text{cos}\,\theta_{\text{Si}}}. \label{eq:ATR_r} 
\end{flalign}

Then, the attenuated total reflectivity ATR\,=\,$r_{\text{Si2}}$/$r_{\text{Siair}}$ is used to calculate amplitude and phase, considering the reference situation of the prism exposed to air ($n_{\text{air}}\,=\,1$) \cite{Nagai2006}. In case of an unknown liquid below the prism, this configuration can help determine the optical constants for angles larger than the critical angle. This method has been applied in several liquids, especially aqueous bioanalyte solutions \cite{Nagai2006, Yada2008, Yada2009a, Max2009, Crompton2012}. \\

\textbf{Under what circumstances are THz measurements typically performed?} \\
In order to avoid the signal amplitude to decay to the noise floor, the interaction path length between THz waves and absorbing medium is reduced within traditional THz sensing approaches. In most cases, the short interaction length is compensated by high analyte concentrations to yield a measurable signal response. For traditional THz sensing techniques, a typical measurement is performed under idealized conditions (enrich analyte concentrations, purify sample matrices, decrease temperature, etc.) and therefore non-physiologically. \\ 

\section{Overview and categorization of biomolecular analytes studied by THz techniques in aqueous environments} 
In the following sections, physiological relevant analytes, e.g., molecules that are a meaningful part of signal cascading, immune response, metabolisms, and synthesis processing, are considered and classified in typical subcategories. To some extent, artificial molecules with biologically relevant properties are also considered. Basically, the study of such candidates at the molecular level can reveal valuable mechanistic information.  In the scope of this article, experiments with organic solvent solutions such as methanol, acetonitrile, and kerosene were excluded in order to focus on bioanalytes in their native aqueous environment. These analytes have been measured using one of the traditional THz measurement techniques (Figure\,\ref{TTF}) or their schematic derivatives.

\begin{figure*}
\centering
\includegraphics[width=\textwidth]{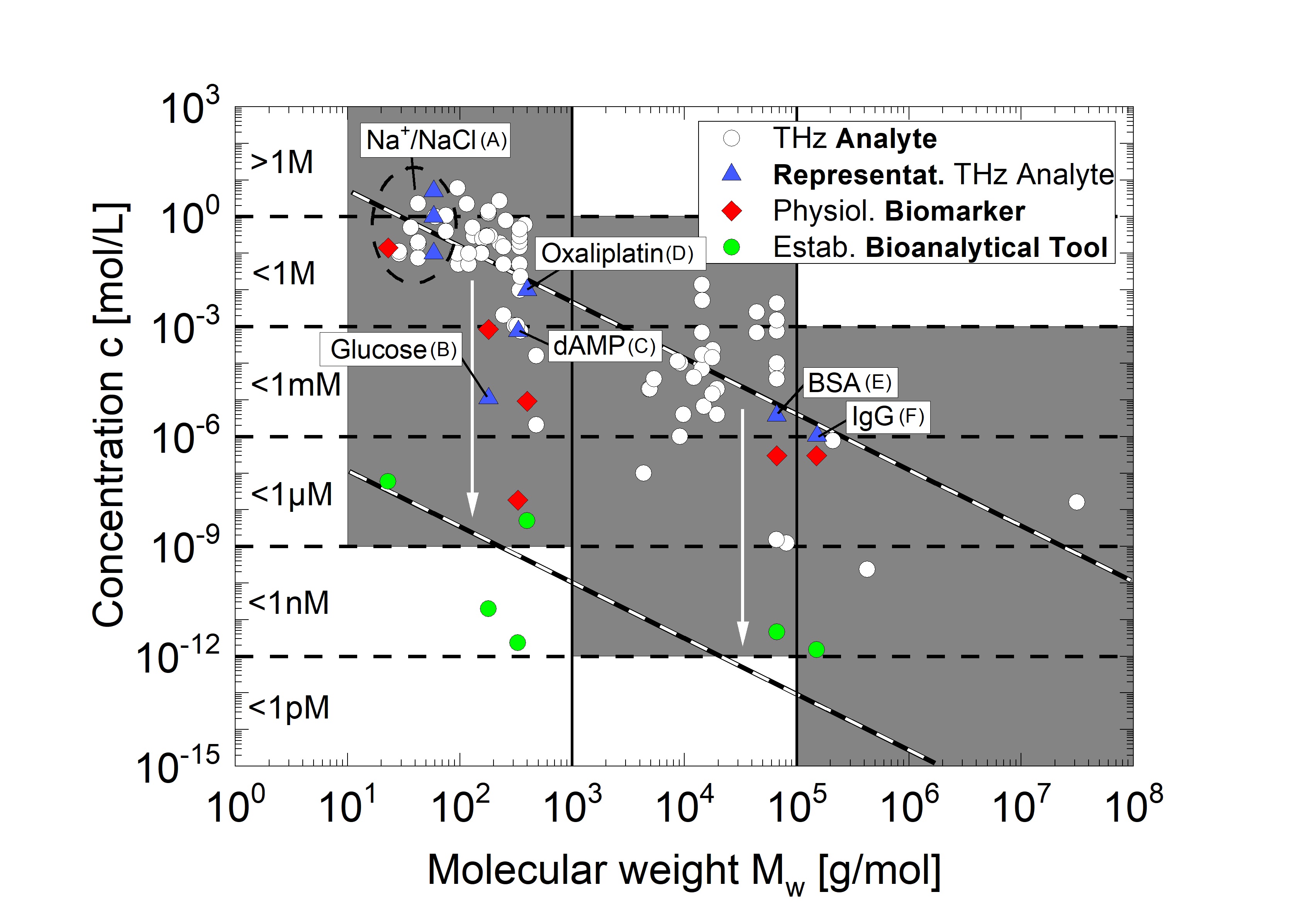}
\caption{Diagram of the reported sensitivities (LDL values) of physiological relevant analytes measured using the traditional THz sensing techniques (white dots). The upper dashed line represents the overall trend of decreasing detection limits with increasing molecular weights. The representative analytes are highlighted (blue triangles). Different concentration levels of the same analyte refer to different reported values from different groups (e.g., NaCl). For comparison, the representative physiological concentration levels are marked (red squares). Typical detection limits are shown for relevant state-of-the-art bioanalytical sensing techniques (green dots). The lower line shows the parallel shifted average THz sensitivity line. The spacing between both lines (indicated by the white arrows) represents the average sensitivity distance between established bioanalytical assays and THz sensing techniques. (An active version of this plot, indicating a citation for every measured point can be accessed online under \cite{HQE} and will provide actualization capabilities for new data.)}
\label{p4}
\end{figure*}

An overall assessment (to the best of our knowledge) of all THz biomolecular sensing experiments performed in aqueous environments to date is presented. For this purpose, in Figure\,\ref{p4}, the measured analyte concentrations that are considered in all reviewed THz studies are depicted as a function of the molecular weight (depicted as white circles). The lower detection limit (LDL) is plotted for the THz experiments, which is defined by the lower concentration level that generates a measurable and distinguishable signal change. For those cases in which analyte concentrations were only reported by weight $\rho$\,[g/L], the molar concentrations $c$\, [mol/L] (equivalent to molarity $c$\, [M]) is calculated to compare all reported values. Furthermore, we depict (patho-)\-physio\-logical concentration levels of representative analyte candidates (red squares) which were taken from reference ranges and indicating typical analyte concentration levels. The LDL values of the selected representative bioanalytical state-of-the-art sensing techniques (green circles) are also depicted in order to illustrate the detection limit performance intercomparison between these methods and the established THz-measurement techniques. All statements derived from this diagram are generally true for all analytes within the ambit of this review, resulting in the conclusion that the selection of representative analytes and reviewed studies in aqueous environments at THz frequencies is consistent. 

For a clear data visibility, we subdivide the analytes in Figure\,\ref{p4} with vertical lines into three informal categories \textit{light} (M$_w\,<\,$10$^3\,$\,g/mol), \textit{medium-weight} (M$_w\,\leq\,$10$^5\,$\,g/mol) and \textit{heavy} (M$_w\,>\,$10$^5\,$\,g/mol) molecules according to their molecular weights. Generally, the molecular weight correlates with the molecular size. Therefore, in the ambit of this review article, it is reasonable to use both expressions synonymously in the following discussion. In addition to the rough weight category, we systematically classify the analytes into one of the following six biomolecular functional categories: (A) The ion group that are considered as the analytes of the lowest complexity. (B) Carbohydrates that are considered as important energy carriers for synthesis processes. (C) Amino acids which act as building blocks for proteins as well as mono- and oligonucleotides. (D) Active agents that are considered to have particular relevance in the development of therapeutic drugs. (E) Light and medium-weight proteins (M$_w\,\leq\,$10$^5\,$\,g/mol) that have multiple active functions in the cell. (F) Large proteins as well as other complex and heavy biomacromolecules (M$_w\,>\,$10$^5\,$\,g/mol).

Some general inferences can be drawn considering the sensitivity diagram in Figure\,\ref{p4}. (i) For light molecules, the concentrations of the THz-investigated analyte range between the mM and M levels. For molecules of lowest complexity such as ions, the physiological concentration levels can be partially detected by using the traditional THz sensing techniques. This is indicated by the red and white symbols within the elliptic dashed line (Figure\,\ref{p4}). (ii) In the case of heavy macromolecules the experimentally accessible concentration levels decrease significantly for higher molecular weights. In comparison to low molecular weight compounds, this difference can be up to ten orders of magnitude lower for compounds with high molecular weight. (iii) Considering medium weight molecules (10$^3$\,g/mol$\,\leq\,$M$_w\,\leq\,10^5$\,g/mol and partially M$_w\,<\,10^3$\,g/mol), such as carbohydrates, nucleotides, active agents, and proteins, the overall trend shows that the detectable concentration levels decrease with increasing molecular weight. (iv) The reported THz concentration levels (white circles) are several orders of magnitude higher (distance blue triangles to red squares) than the physiological concentration levels (red squares). (v) In general, state-of-the-art bioanalytical techniques (green circles) are also several orders of magnitude more sensitive than traditional THz sensing techniques (blue triangles). In general, deviations from this standard in (iv) are discussed below in the corresponding analyte group. (vi) For both, THz sensing and state-of-the-art bionalytical methods, the overall trend of decreasing detection limits with increasing molecular weight is indicated by black dashed lines. The analysis of this wide collection of measurements indicates together with the fitting of the lines that the lower detection limit scales with \textcolor{black}{LDL$\,\propto\,$M$_w^{-1.53\pm0.13}$}, which is a significantly smaller exponent than \mbox{m\,$\approx$\,-1} and indicates a super linear relationship between LDL and molecular size or mass. At this point, the determination of an exact relationship between LDL and molecular weight is difficult because of the extremely variational nature of the items in the analyte library. Still, an exponent of \textcolor{black}{-1.53} indicates the capability of THz approaches for sensing more delocalized molecular modes that increasingly affect analytes and their environments with growing molecular size. 

Based on the diagram in Figure\,\ref{p4}, the detection and characterization potential of biomolecules in their native aqueous environments by using THz spectroscopy is discussed in more detail in the following sections. At least one representative molecule is presented for each of the six groups for discussion of (patho-)\-physio\-logical relevance and concentrations in more detail. Furthermore, the sensitivity obtained by using THz-based techniques is compared to the LDL value of  established bioanalytical systems for each corresponding analyte type. Subsequently, the prospective THz detection and characterization potential is evaluated with regard to the analyte group in question.

\subsection{Ions}
\begin{figure*}
\centering
\includegraphics[width=0.70\textwidth]{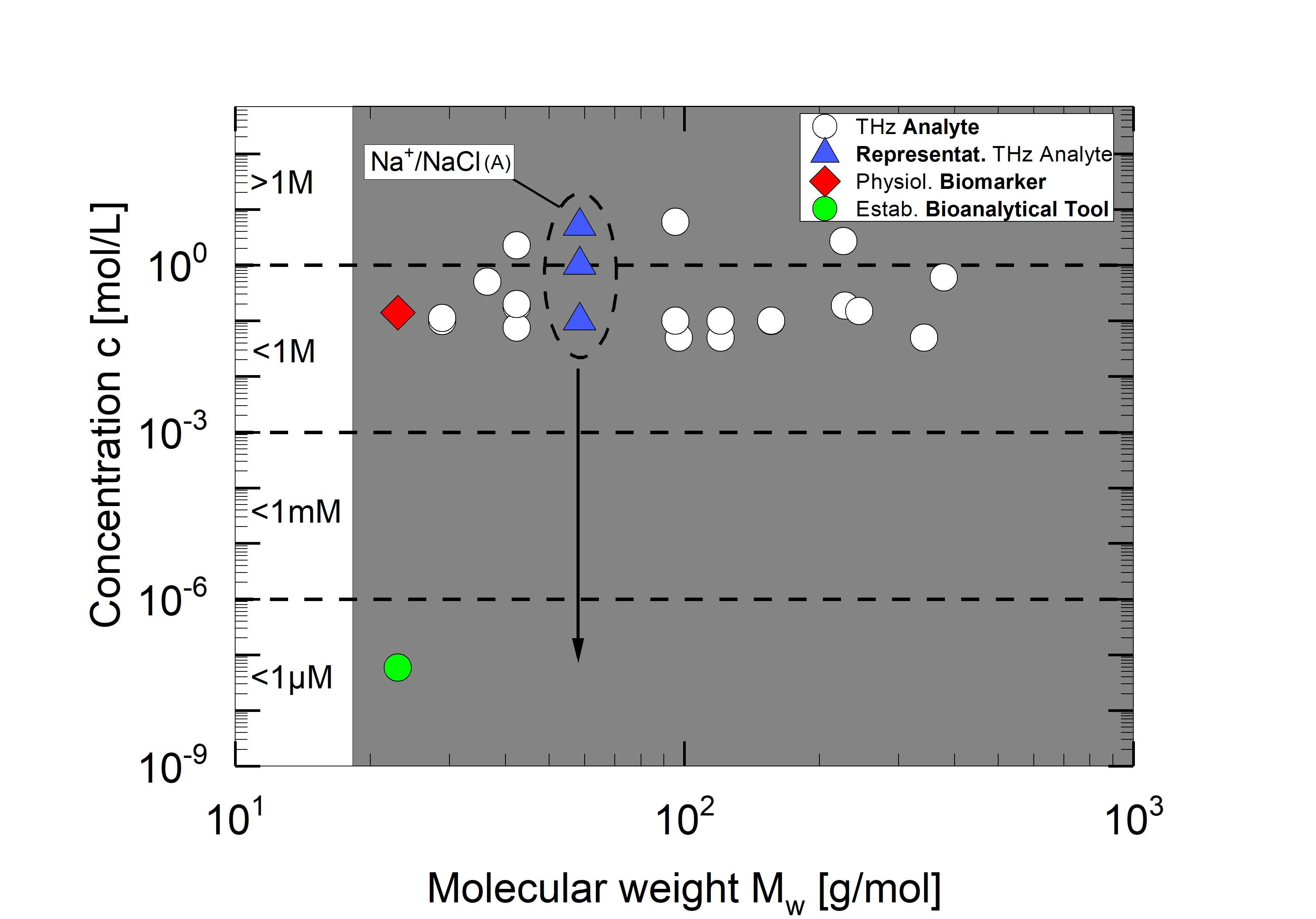}
\caption{\textcolor{black}{Diagram of the reported LDL values of ions using traditional THz techniques (white dots), representative analytes (blue triangles), representative physiological concentration levels (red squares) and relevant state-of-the-art bioanalytical techniques (green dots).}}
\label{LDL_Ions}
\end{figure*}

For the THz-based investigation of aqueous media, ions are suitable analytes because they can significantly change the status of charge inside the liquid. This has a direct impact on the dynamics of the interaction of water molecules, which leads to a large THz response \cite{Smiechowski2013}. \textcolor{black}{An overview of the existing investigations on ions is shown in Figure\,\ref{LDL_Ions} (presenting a sub-selection of datapoints for ions from Figure\,\ref{p4}).}\\

\subsubsection{Subject of THz studies}
It has been experimentally shown that the addition of salt ions to aqueous solutions can significantly change the dielectric response. The dielectric constants can subsequently be calculated from the recorded spectrum \cite{Xu2006b}. Generally, for the theoretical description of the frequency-dependent dielectric function the permittivity is calculated from multi Debye modeling approaches. From the modeled dielectric response, the dynamic parameters of the environment (H$_2$O molecules in the ambit of this review) that are adjacent to the ionic analyte can be estimated \cite{Asaki2002, Koeberg2007, Post2013, Xu2007, Jepsen2010, Qiao2012, Wu2013, Sebastiani2020}. Beyond that, attempts have been made to categorize ions into structure-strengthening and structure-weakening compounds according to systemized THz spectroscopic data \cite{Kaun2005}. It is observed that the interpretation of the dielectric response of comparatively simple solutions under test is not trivial. A major contribution to the overall dielectric response of the salt modes rather than the water network modes has been indicated by an experiment investigating the impact of alkali halide salts on water network dynamics \cite{Schmidt2009}. A direct classification of chaotropic and kosmotropic analytes solely based on THz spectroscopy data is difficult to realize according to this finding. In addition, detailed models are developed to support the determination of the origin of the changed dynamics. For example, the dissection of LaBr$_3$ and LaCl$_3$ broadband THz spectra into anion, cation, and ion pair contributions was demonstrated using a chemical equilibrium model \cite{Sharma2013}. \\

\subsubsection{(Patho-)\-physio\-logical relevance and concentration level of Na$^+$ ions}
In this review article, Na$^+$ ions are selected as representative analytes. For example, they are a significant constituent part of ion transporting systems, which is important for maintaining or creating concentration gradients, electrical potential, or regulating the cellular transfer of substances through the cell membrane \cite{Morth2011}. An unbalanced sodium concentration level, e.g., hyponatremia ($c\,<\,$136\,mM serum level) \cite{Adrogue2000}, can indicate a number of life-threatening diseases, such as chronic renal failure \cite{Kaji1987} and cancer \cite{AbuZeinah2015}. Generally, in healthy bodies the physiological Na$^+$ concentration level depends on the location. In the extracellular space the reference range is approximately given in the range between $c$\,=\,120\,mM and 150\,mM \cite{Madelin2014,Madelin2015,TAGUCHI2018}, whereas the intracellular concentration is significantly lower ($c$\,=\,8{--}15\,mM)\cite{Stryer2013,Rottman1992,Madelin2015}. \\

\subsubsection{THz spectroscopy versus established bioanalytical measurement system for Na$^+$ ions}
The relevance of Na$^+$ ions as a biomarker in clinical diagnostics can be considered high. In fact, using traditional THz sensing techniques, the physiological concentrations of Na$^+$ ions can be detected. Experimentally detected concentrations have been reported in the 100\,mM range \cite{Schmidt2009, Qiao2012}. However, the comparison between THz-detected Na$^+$ sensitivities in \textcolor{black}{Figure\,\ref{LDL_Ions}} and a typical polymer membrane ion-selective electrode sensor shows a sensitivity offset of six orders of magnitude. In clinical diagnosis setups, a high sensitivity of any applied measurement technique is crucial as concentration levels should be detectable nit only in but also below an analytes reference range. Ion-selective electrodes are very effective sensors that have shown a detection limit of $c$\,=\,59\,nM for Na$^+$ ions \cite{Malon2006}. Such sensitivities have not been demonstrated with THz sensing tools, yet. \\

\subsubsection{Perspectives of THz sensing of ions}
In summary, it is interesting to raise the question if there is a demand for alternative THz-based techniques to detect biochemically relevant ions at physiologically relevant concentrations. Considering the above-explained hyponatremia, the determination of the medical cause supported by differential diagnostic method algorithms is a major challenge for the medical staff. In order to achieve successful treatment by the choice of an effective therapeutic approach, it is important to be able to make well-founded decisions at an early stage already. As a matter of fact, established detection techniques work well for the complete determination of Na$^+$ concentrations.

The THz-based measurement technique is an ensemble averaging method \cite{Koeberg2007}. Since it introduces non-specificity, the THz-based detection of Na$^+$ ions can become difficult. It is not trivial to assign the origin of a multimodal-modified dielectric response from the THz data alone because of the non-specificity. Therefore, the capability of the THz ion detection is clearly restricted to applications that are based on high ionic contrasts. Masson et al. \cite{Masson2006} reported a potential THz application for imaging. They introduced a near-field ion contrast microscopy scheme for imaging neuronal tubes obtained from worms and preserved in physiological solutions. However, these experiments have not been demonstrated \textit{in vivo} yet. It would be very interesting to see this technique working under more complex physiological conditions. 

In the near future, we do not expect Na$^+$ ions to be used as a biomarker in THz applications for clinical diagnostics directly, unless solutions to the non-specificity of the existing THz excitation and probing methods are provided.

\subsection{Carbohydrates}
Carbohydrates are one example for molecular energy carriers. Especially pentose- and hexose-based compounds are important candidates as indicated by their large number of derivatives. For example, glycoproteins which are conjugates of oligosaccharides and membrane proteins are important constituents of enormous structural diversity. In addition, they play an important role in cell-to-cell communication by creation of cell-\textcolor{black}{selective} surfaces. Thus, they can be considered as an information storage medium. Glycoproteins can also function as antigen determining groups. Probably, the most well-known carbohydrate derivative is 2-deoxy-D-ribose which is one of the three structural elements of DNA and DNA nucleotides (cf. \cite{Stryer2013, Bannwarth2013}).

\begin{figure*}
\centering
\includegraphics[width=0.70\textwidth]{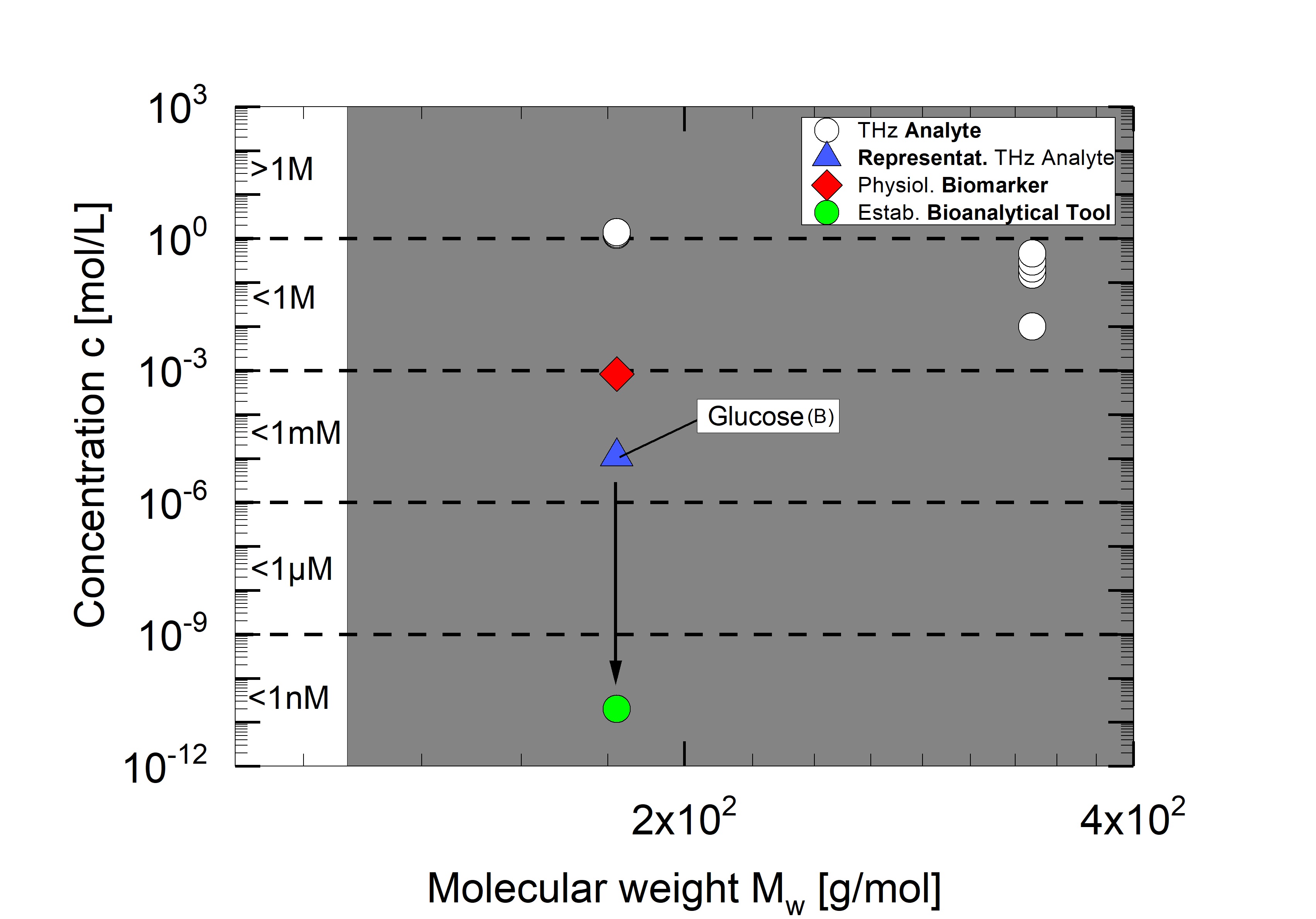}
\caption{\textcolor{black}{Diagram of the reported LDL values of carbohydrates using traditional THz techniques (white dots), representative analytes (blue triangles), representative physiological concentration levels (red squares) and relevant state-of-the-art bioanalytical techniques (green dots).}}
\label{LDL_carbohydrates}
\end{figure*}

Scientific questions, such as "how are carbohydrates linked to proteins on a functional level?" and "how are they recognized in a complex biochemically relevant medium?", took part in the development of glycobiology as an autonomous field of research. In the analytical context of glycobiology, the THz community has intensely examined carbohydrate derivatives in aqueous environments. \textcolor{black}{An overview of the existing investigations on carbohydrates is shown in Figure\,\ref{LDL_carbohydrates}.}\\

\subsubsection{Subject of THz studies}
For the successful spectroscopic detection of carbohydrates, experimental investigations proved the feasibility of traditional THz sensing schemes. To detect and characterize glucose and sucrose, both transmission and reflection geometries are used according to their optical constants \cite{Cheng2008, Nagai2006, Jepsen2007, Serita2019, Song2018}. The characterization of solvated lactose hydration shells \cite{Heugen2006} and the estimation of hydration states of sucrose as well as trehalose have also been demonstrated \cite{Arikawa2008}. \\

\subsubsection{(Patho-)\-physio\-logical relevance and concentration level of glucose}
In this review article, for the representative physiologically relevant analyte, glucose was chosen from the carbohydrate group. In the carbohydrate metabolism of an organism, glucose plays a crucial role. It is either supplied with food or synthesized by gluconeogenesis in the liver \cite{Stryer2013}. The reference ranges of glucose can differ from one another depending on the body fluid of interest (e.g., plasma, whole blood, and urine). On average, the typical concentrations of glucose range in the low mM levels \cite{Gressner2013}. An unstable glucose metabolism can lead to glycemia \cite{Cryer2003}. For example, hyperglycemia (c$\,>\,6.1$\,mM in whole blood) indicates diabetes mellitus, which is an endemic disease that affects millions of people \cite{Gressner2013, Harper2010}. High glucose concentration levels in the non-diabetic regime (c$\,>\,$5.5\,mM) indicate an increased risk of heart diseases in non-diabetic subjects as shown by \cite{Laakso1999}. \\

\subsubsection{THz spectroscopy versus established bioanalytical measurement systems for glucose}
Since the early 1960s, an ongoing research, development, and improvement program is being carried out in numerous electrochemical sensor schemes that are suitable for glucose detection \cite{Harper2010, Wang2008, Tian2014}. Based on electrochemical detection of H$_2$O$_2$, a widespread scheme is generated during the glucose oxidase method. Generally, concentrations in the nM to the lower $\mu$M range can be accessed with electrochemical biosensors \cite{Harper2010, Chen2013}. 

In addition, fluorometric assays are available in which glucose concentrations of 20\,pM can be detected in blood, plasma, or urine \cite{sigmaGlucose}. \textcolor{black}{As it can be seen from Figure\,\ref{LDL_carbohydrates}, t}he glucose concentration detection limit measured using electrochemical sensors or fluorometric assay kits is six orders of magnitude better than the most sensitive THz spectroscopy approach reported by Serita et al. (c\,=\,11.1\,$\mu$M, \cite{Serita2019}). Nevertheless, their introduced non-linear optical crystal (NLOC)-based THz-microfluidic chip with asymmetric metamaterials has an at least two orders of magnitude lower detection limit compared to previous published THz results. With this technique it is therefore already possible to reliably detect mM glucose concentrations as they occur in blood. \\

\subsubsection{Perspectives of THz sensing of carbohydrates}
What is the potential for THz applications in clinical diagnostics for bioanalytical applications linked to carbohydrates? Basically, a biosensor is characterized by multiple properties, like detection limit, selectivity, robustness, long-term stability, performance, and response behavior but also application specific aspects such as invasiveness, analytic complexity or temporal response. Electrochemical state-of-the art biosensor techniques demonstrate good detection limits considering the sensitivity issue. However, before achieving a highly stable and reliable operation by monitoring the glucose level, there are still many challenges to overcome. Toward the suitability for daily life healthcare applications, moment-to-moment monitoring is considered to be a crucial step (cf. \cite{Wang2008}). The determined glucose concentration levels may also help in monitoring the decreased lactase activities that indicate lactose intolerance. THz techniques seem to be very interesting to this regard as no direct contact between sensor and medium is required. The recently published work of Serita et al. \cite{Serita2019} is a promising approach for detecting carbohydrates at physiological concentrations. However, the non-specificity is potentially problematic and hinders the straightforward implementation as a tool to detect carbohydrates in complex physiologic liquids that is useful for clinical diagnostics or healthcare applications. Current developments using advanced numerical procedures indicate the feasibility to overcome such obstacles \cite{martin2016vivo}, which is extremely attractive with regard to an extension of THz analyses by a \textcolor{black}{selective} detection capability in order to enable an actual use for widespread clinical diagnosis.

Apart from the clinical THz detection of carbohydrates, a huge effort has been put in understanding functional mechanisms at the molecular level in more detail. For the investigation of carbohydrates, it has become evident that THz radiation can be a strong characterization instrument to measure the size of hydration shells of molecules like lactose. Regarding bioprotection, which is mediated by constrained water shells, the important functionalities of the solvation shell have been confirmed by comparing experimental data and theoretical considerations \cite{Heugen2006, Massari2005}. Such THz investigations can be precious tools to understand and determine the impact of solvation dynamics on molecular functionality, not only for carbohydrates but also for other physiologically relevant compounds. These interesting results might be a basis for future experiments since more specific analysis will be required to discern solvation in significantly more complex molecular embeddings closer to physiological conditions, yet.

\subsection{Amino acids and nucleotides}
 The members of the amino acid and nucleotide group are of particular physiological relevance, since they represent the building blocks of more complex molecular units. Chemically speaking in a simplified form, nucleotides are monomers for the resulting polymer called DNA on whose sequence a beings genome depends. There are closely related derivatives forming RNA, both of which get translated into proteins in a living cell in the last consequence \cite{Stryer2013, Lottspeich1998}. Proteins on the other hand are composed of amino acid precursor units \cite{Stryer2013}.

\begin{figure*}
\centering
\includegraphics[width=0.70\textwidth]{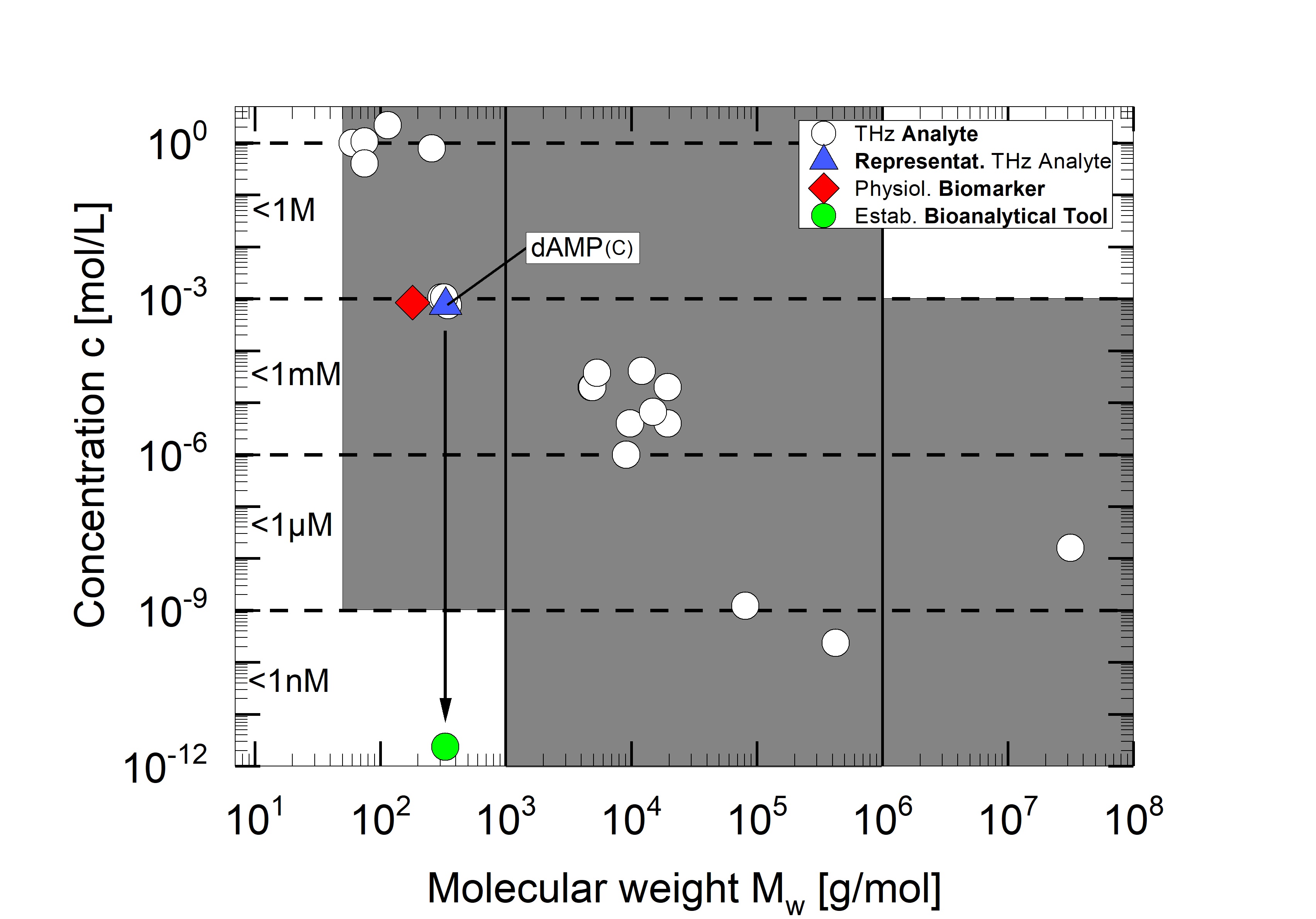}
\caption{\textcolor{black}{Diagram of the reported LDL values of amino acids and nucleotides using traditional THz techniques (white dots), representative analytes (blue triangles), representative physiological concentration levels (red squares) and relevant state-of-the-art bioanalytical techniques (green dots).}}
\label{LDL_acids}
\end{figure*}

A large number of vibrational modes can be excited by THz waves within and between these molecules \cite{Plusquellic2007}. In addition, because of their atomic and structural composition, these biomolecules contain a significant amount of charge as soon as they are dissolved in water \cite{Lottspeich1998}. There is negative charge in case of nucleotides due to phosphate in the backbone, while the charge state for amino acids depends on the functional groups and on the pH of the medium \cite{Lottspeich1998}. Therefore, amino acids and nucleotides are considered attractive targets for THz-based analysis in aqueous media because of the high interaction potential with THz waves. \textcolor{black}{An overview of the existing investigations on amino acids and nucleotides is shown in Figure\,\ref{LDL_acids}.}\\

\subsubsection{Subject of THz studies}
Consequently, molecules such as amino acids (and urea as a metabolite), nucleosides, nucleotides, and nucleic acids are investigated by the THz community. The characteristics of these molecules, such as temperature-dependent absorption spectra \cite{Funkner2012, Cooksey2009}, the dielectric response of molecular solutions such as completely solvated mono- and oligonucleotides \cite{Glancy2010, Arora2012, Sun2014, Tang2018, Ogawa2009, tang2020, tang2020a, Shih2018, Jeong2020}, modifications of the molecular structure and spectral signatures as a function of both the polarization state of incident radiation \cite{Globus2006} and the hydration level \cite{Lvovska2010}, have been reported. \\

\subsubsection{(Patho-)physiological relevance and concentration level of nucleotides}
We primarily focused on deoxyadenosine monophosphate (dAMP) among the group of deoxyribonucleotides. Since concentration levels of the analyte range significantly below saturation, the investigation of deoxyribonucleotides in aqueous environment carried out by Glancy et al. \cite{Glancy2010} is particularly interesting. The structurally closely related cyclic adenosine monophosphate (cAMP) also has a \mbox{(patho-)}\-physio\-logical relevance. Pseudohypoparathyroidism is indicated by anomalies in the reference range, which is generally associated with tumor diseases \cite{Naide1968}. The physiological concentrations range around the lower nM levels (18.3{--}45.5\,nmol/L) \cite{Gressner2013} considering glomerular-filtrated cAMP. \\

\subsubsection{THz spectroscopy versus established bioanalytical measurement system for nucleotides}
The comparison in \textcolor{black}{Figure\,\ref{LDL_acids}} reveals that the LDL value of an exemplary ELISA \cite{Enzo} for the detection of cAMP is by nine orders of magnitude better than the LDL value detected with a THz-TDS-Spectrometer of dAMP in solution \cite{Glancy2010}. \\

\subsubsection{Perspectives of THz sensing of nucleic acids and nucleotides}
There is a large difference in the sensitivity of THz sensing of nucleic acids. Arora et al. for example have reported significantly enhanced LDL values for THz-supported detection of polymerase chain reaction (PCR) amplified DNA in aqueous media \cite{Arora2012}. They have shown sensitivities in the nM-pM range. This LDL value represents a record-breaking number independently of its molecular when compared to all other experiments that are taken into account in the scope of this review. We consider two aspects here. Firstly, the high sensitivity follows a fragile balanced differential measurement technique \cite{Bergner2005} that has a capability of recognizing even smallest differences in a sample by removing the water absorption, and thereby significantly reducing the background signal. This differential technique can detect spectacularly low LDL values compared with standard THz measurement techniques. Secondly, for detection of the nucleic acid sequence, there seems to be an ongoing demand for the development of label-free and sensitive DNA detection strategies in order to reduce preparation steps and time besides avoiding costly chemical procedures like fluorescent labeling, for instance. This perception is widely accepted in the THz community. However, colorimetry-based analytical methods enable single-base discrimination \cite{Li2005}. The detection of DNA presence in a solution plays an insignificant role in nucleic acid analytics. Therefore, many researchers focus on further exploration of technical potential to detect single-base mismatches with biochip arrays. Promising results have been obtained in dried aqueous samples since the early analysis of label-free THz detection of DNA samples and single-base mutations \cite{Nagel2002, Bolivar2002, HaringBolivar2004}. By providing the necessary THz sensitivity and \textcolor{black}{selectivity} to measure DNA sequences from complete cell lysates at physiological concentrations, and obliviating the need for PCR amplification, another milestone was achieved recently \cite{weisenstein2020ultrasensitive}. Precisely, if such THz-based sensing is developed as a powerful alternative for bioanalytical standard assays and techniques, its detection and characterization potential in a standard aqueous environment needs to be clearly evaluated against established techniques. \\

\subsection{Active agents}
\begin{figure*}
\centering
\includegraphics[width=0.70\textwidth]{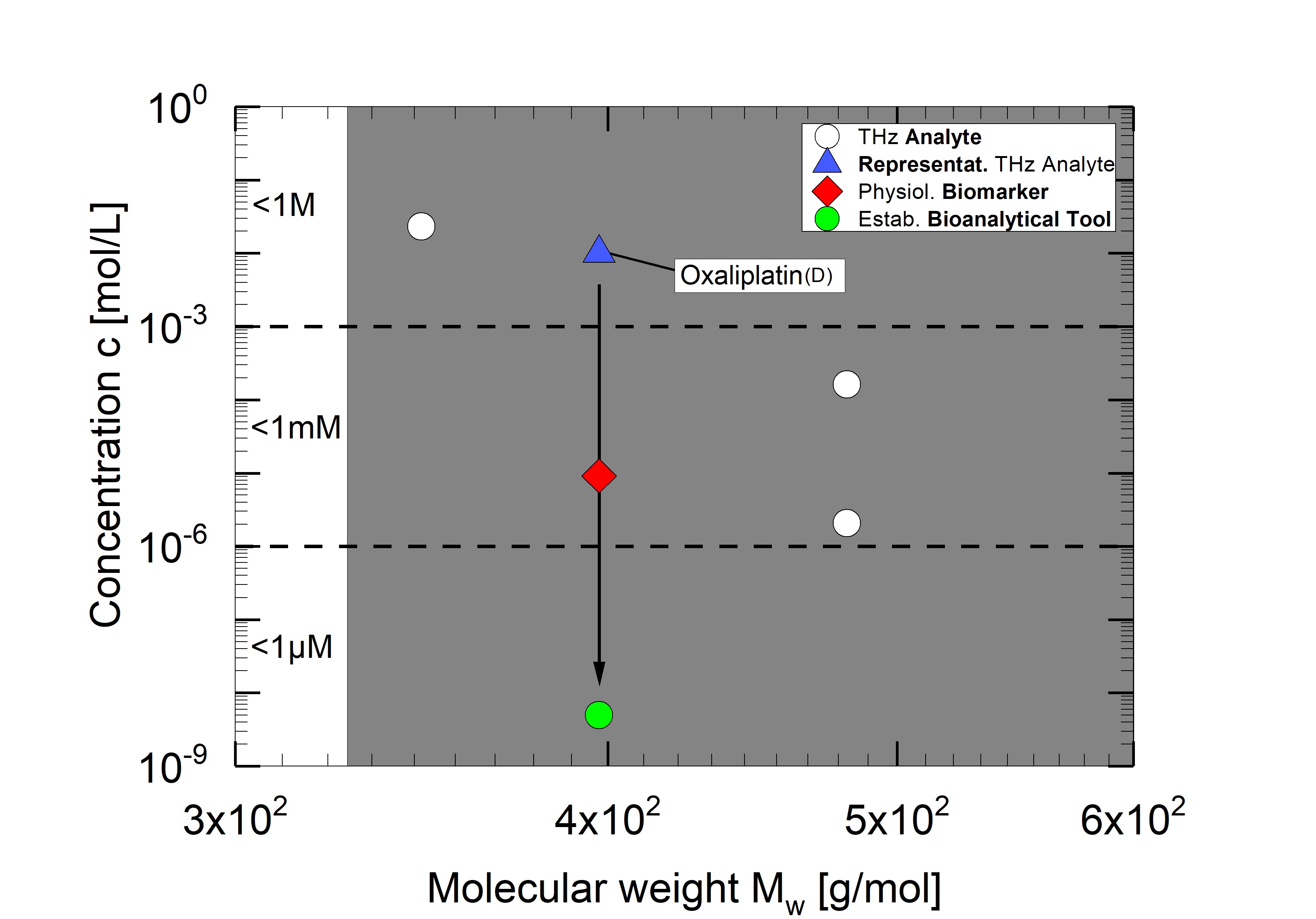}
\caption{\textcolor{black}{Diagram of the reported LDL values of active agents using traditional THz techniques (white dots), representative analytes (blue triangles), representative physiological concentration levels (red squares) and relevant state-of-the-art bioanalytical techniques (green dots).}}
\label{LDL_ActiveAgents}
\end{figure*}

Since active agents can have different targets, such as macromolecular nucleic acids or proteins (e.g., enzymes), these compounds play a significant role in the development of pharmaceutical drugs. The significance of active agents in therapeutic treatment of diseases is very high, accordingly. In particular, a strong interest exists in revealing functional mechanisms at the molecular level, which are not entirely understood in many cases \cite{Karimi2014}. In this regard, a contribution to a more comprehensive mechanistic knowledge, for example by enhanced screening methodologies might improve the efficiency of pharmaceutical drug development. \textcolor{black}{An overview of the existing investigations on active agents is shown in Figure\,\ref{LDL_ActiveAgents}.}\\

\subsubsection{Subject of THz studies}
First THz spectroscopy experiments have been carried out using a drug that is used to reduce blood pressure called nifedipine. These experiments have shown that the structural composition of crystallite depends on the water content in the analytic environment \cite{Takebe2013}. Moreover, THz investigations focused on characterizing the reaction dynamics between DNA and oxaliplatin, which is an anticancer substance \cite{Wu2012}. THz applications in this field emerge but are not widely addressed, yet. In the field of the frequently used antibiotics, the two active ingredients doxycycline hydrochloride (DCH) and tetracycline hydrochloride (TCH) have been studied using metallic mesh based reflection terahertz spectroscopy and attenuated total reflection terahertz spectroscopy, respectively\cite{Wang2020, Qin2017}.\\ 

\subsubsection{(Patho-)physiological relevance and concentration level of oxaliplatin}
For this section, we have selected oxaliplatin as a representative active agent because of its high pathophysiological importance being an anticancer drug. In the chemotherapeutical treatment of cancer, the investigation of platinum-based antitumor compounds such as oxaliplatin attracts a high degree of attention. The metabolites of this substance can be activated in aqueous environments. They subsequently bind to DNA, thus modifying the macromolecule by intra- and intermolecular strand connections. Cell activities including DNA (e.g. DNA replication and transcription) are effectively inhibited by these modifications (cf.\,\cite{Crisafuli2012}). The concentration of oxaliplatin has been reported in the lower \,$\mu$M range during a two hour infusion as shown by pharmacokinetic investigations. Ip et al. reported a mean maximum value of $c$\,=\,9.2\,$\pm$\,1.4\,$\mu$M \cite{Ip2008}. \\

\subsubsection{THz spectroscopy versus established bioanalytical measurement system for oxaliplatin}
A high-performance liquid chromatography inductively coupled plasma mass spectrometry (HPLC-ICP-MS) assay is used for detection of oxaliplatin levels in the plasma of the patient \cite{Ip2008}. The LDL value for this HPLC-ICP-MS assay ($c\,=\,5$\,nM) is six orders of magnitude smaller than the sensitivity obtained by using a THz spectroscopy sensing platform as demonstrated by Wu et al. ($c\,=\,10$\,mM) \cite{Wu2012}, \textcolor{black}{see Figure\,\ref{LDL_ActiveAgents}}. However, it should be noted that an HPLC-ICP-MS assay is suitable for medical analyses to a limited extent, as sample preparation is very time-consuming and expensive. \\

\subsubsection{Perspective of THz sensing of oxaliplatin}
Active agents such as oxaliplatin have a high pharmaceutical relevance. For medical research as well as for treatment applications, their reliable, robust, and fast detection in body fluids at a molecular level is of very high importance. THz sensing considered as a potential scheme for real-time and \textit{in vivo} monitoring of the concentration levels of an active agent during chemotherapeutical treatment appears very attractive for adjusting optimally matched drug profiles during infusions. Here, the emphasis is entirely on the \textcolor{black}{selective} target molecule detection, while a possible \textit{in vivo}, real-time, and label-free THz application would represent a huge impact scenario. In this regard, the THz community needs to develop strategies to overcome limitations imposed by non-specificity issues under physiological conditions.

Moreover, there is a great deal of interest in decoding reaction mechanisms at the molecular level. One of the biggest challenges in the scientific world is the characterization of reaction mechanisms and dynamics at the sub-molecular level in real time. In this regard, a better understanding of active agents  would also improve the pharmaceutical drug development and optimization. With first valuable THz sensing experiments, initial steps are already made in this direction. \\

\subsection{Proteins}
\begin{figure*}
\centering
\includegraphics[width=0.70\textwidth]{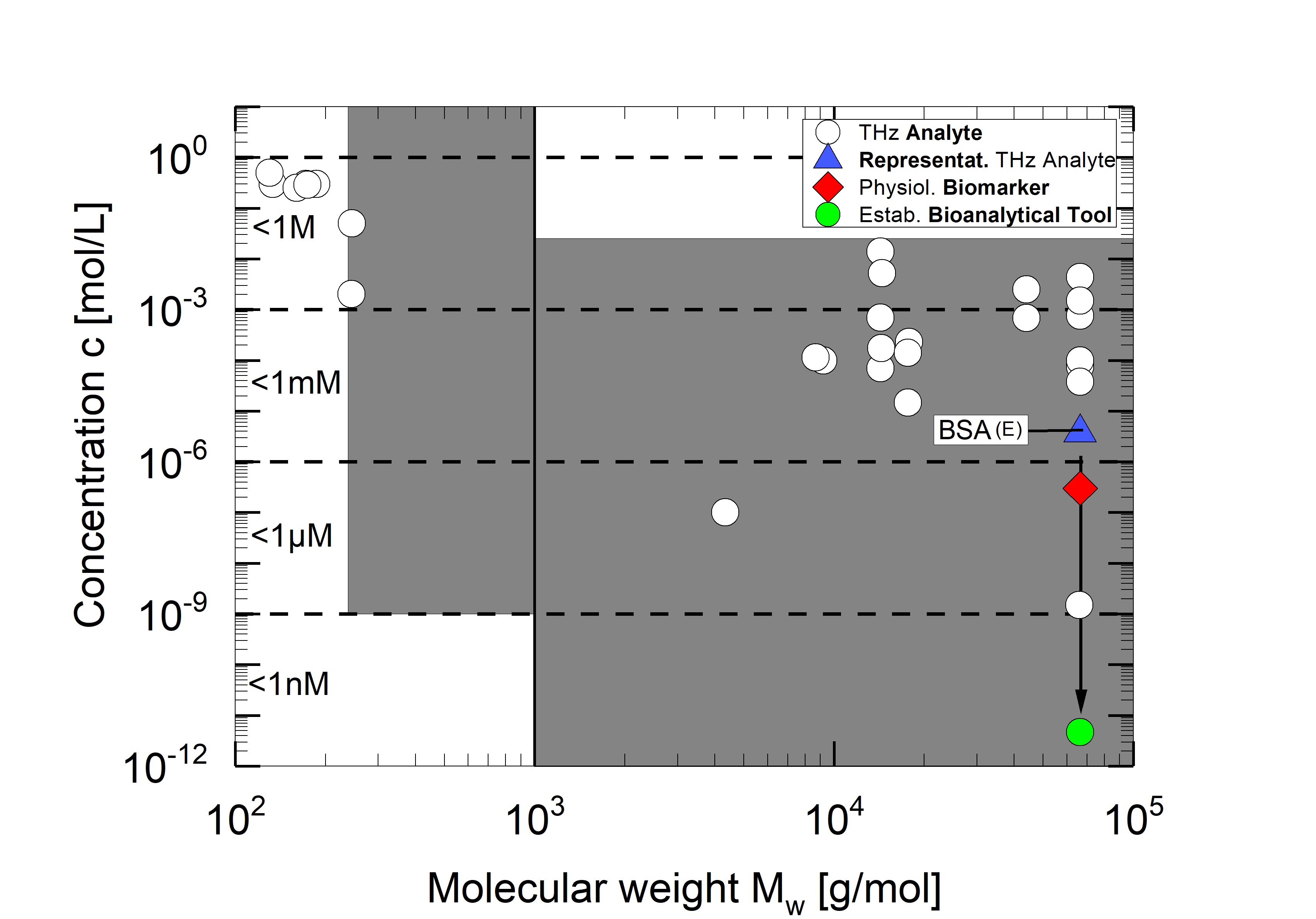}
\caption{\textcolor{black}{Diagram of the reported LDL values of proteins using traditional THz techniques (white dots), representative analytes (blue triangles), representative physiological concentration levels (red squares) and relevant state-of-the-art bioanalytical techniques (green dots).}}
\label{LDL_Proteins}
\end{figure*}

In the biochemical context of a living organism, proteins play the role of the functional units. In order to explain pathogenic factors or disease progression, for instance, a lot of effort is put in understanding the complex biochemical processes at the molecular level, which are generally found closely related to protein dysfunctions \cite{Chiti2006, DalleDonne2006}. Proteins also play a major role in the development of pharmaceutical drugs because most drug targets are built up from protein structures \cite{Overington2006}. Additionally, in clinical diagnostics proteins have a very high impact because because the concentration levels of relevant biomarkers, when outside of pathophysiological reference ranges, often indicate physiological malfunction \cite{Gressner2013}. There is still an ongoing requirement for precise, robust, sensitive, and fast alternative measurement techniques to detect biomarkers, despite of the wide range of established bioanalytic tools. This is essential because of the complexity of the sample purification and preparation procedures of existing approaches. The evaluation of the THz sensing potential for label-free protein detection deserves special attention, therefore. \textcolor{black}{An overview of the existing investigations on proteins is shown in Figure\,\ref{LDL_Proteins}.}\\

\subsubsection{Subject of THz studies}
The globular proteins human serum albumin (HSA) \cite{Luong2011} and bovine serum albumin (BSA) \cite{Xu2009, George2008, Dinca2010, Laurette2012, Kitagawa2006, Oleksandr2015, Sczech2014, Zhang2018}, the enzyme lysozyme \cite{Xu2006a, Chen2007, Woods2010, Laurette2012, Oleksandr2015}, the prosthetic group biotin \cite{Baragwanath2010}, the oxygen-carrying muscle protein myoglobin \cite{Zhang2004, Murakami2012, Oleksandr2015, Cheng2020}, the biotin-binding protein avidin \cite{Zhou2020}, the antifreeze protein DAFP-1 \cite{Meister2013}, the photoactive yellow protein \cite{George2013}, alanine-rich peptides \cite{Ding2010}, the peptide amyloid beta \cite{Heo2020}, and the artificial peptide sequences P-NIPAAm \cite{Naito2012, shi2015} have been investigated with traditional THz techniques. Absorption spectra and optical constants have been recorded and determined considering the requirement for reduced interaction between THz fields and absorbing aqueous media \cite{Baragwanath2010, George2008, Dinca2010, Luong2011, George2013, Chen2007, Murakami2012, Kitagawa2006, Naito2011}. Moreover, the dynamics of collective modes \cite{Xu2006a} or temperature-dependent structural changes \cite{Naito2012} have been investigated. In addition, a lot of scientific work is dealing with the problems of quantitative analysis and the parameters that affect hydration and solvation dynamics \cite{Woods2010, Zhang2004, Meister2013, Ebbinghaus2007, Ding2010, Born2009, Xu2009, Laurette2012}. Furthermore, the energy flow from dialanine molecules into the aqueous environment has been studied in the infrared and far-infrared region, respectively \cite{Niehues2012}. \\

\subsubsection{(Patho-)physiological relevance and concentration level of serum albumin}
Serum albumin is the representative candidate that has been selected for this bioanalyte group. It is a large molecule with a molecular weight of 66400\,g/mol and has numerous functionalities in living organisms \cite{Evans2002}. For example, it helps in maintaining the colloid-osmotic pressure, which is responsible for regulating the water distribution between inter- and extravascular space \cite{Prinsen2004}. 
Secondly, albumin is considered a highly charged molecule in blood at physiological pH\,=\,7.2, and thus enables the transport \cite{Merlot2014} of substances with lower solubility. Renal disorder is indicated for physiological densities with values $\rho\,>\,20\,$mg/L ($c\,\approx\,0.3\,\mu$M). An increase in the glomerular filtration of serum albumin has been observed in case of albuminuria \cite{Gressner2013}.\\

\subsubsection{THz spectroscopy versus established bioanalytical measurement system for serum albumin}
For the detection of serum albumin with high sensitivity, a colorimetric ELISA with lower detection limits in the pM range \cite{abcamHSA} can be employed. Alternatively, in order to obtain higher detection limits as well, test stripes with a detection limit of approximately $\rho\,=\,20\,$mg/L ($c\,\approx\,0.3\,\mu$M) can be used. This helps in determining the concentrations of serum albumin in urine (\textcolor{black}{Figure\,\ref{LDL_Proteins}}, $c_{\text{urine}}$\,=\,0.31\,$\mu$M) \cite{Gressner2013}. These higher densities have not been demonstrated with THz sensing techniques yet. However, considerably higher concentration levels of $\rho\,=\,35\,-\,50\,$g/L ($c\,\approx\,0.53\,-\,0.75$\,mM) \cite{peters1995} are accessible with traditional THz sensing techniques ($c_{\text{LDL}}\,=$\,3.8\,$\mu$M) \cite{Zhang2018} as can be seen from the comparison in \textcolor{black}{Figure\,\ref{LDL_Proteins}}. Basically, at physiologically relevant concentrations, this helps in the THz investigation of complex functional molecules. \\

\subsubsection{Perspective of THz sensing of proteins}
The high degree of non-specific THz excitation is still a big issue and the \textcolor{black}{selective} discrimination under physiological conditions is extremely difficult in application-orientated diagnostic detection. Accordingly, for an alternative label-free detection strategy that is suitable for \textit{in vivo} applications, the potential of THz sensing is currently not yet fulfilled; the community needs to find out other scenarios that can be competitively applicable for THz characterization. The term proteomics exists since late 1970s, which summarizes the attempt to detect all expressed proteins under defined conditions at a particular time with bioanalytical methods \cite{Pandey2000}. This approach is based on the idea of identifying and analyzing the full proteome (e.g., of a cell). Ideally, specific expression profiles of gene products are associated with diseases. In reality, of course, this approach has several limitations that are indicated by restrictions related to the measurement techniques or by the high complexity of the medium under physiological conditions. A great amount of analytical effort is required for protein characterization irrespective of type of the bioanalytical tool employed.

Besides proteome characterization, there are well-known approaches for analyzing protein-protein interactions in the field of proteomics. For example, highly selective affinity assays or high-throughput suitable yeast 2-hybrid systems can indicate protein interaction by the transcription of reporter genes \cite{Pandey2000}. Still, such analyses are limited to the detection of binding events only. High sensitivity detection principles in combination with very low detection limits are the best characteristics of state-of-the-art bioanalytical tools. However, it is hardly possible to make substantial statements regarding dynamical processes using such assays. On the contrary, THz energies provide the opportunity to excite and record dynamical processes at the molecular level. This is one of the obvious conditions that could preferentially be occupied by THz sensing. Note that THz-excited delocalized collective modes of complex molecules also enable potential access to the analyte environment. In order to reveal mediating properties, functionalities or even molecular malfunctions, the highest characterization potential for THz sensing of proteins in aqueous solutions is possibly to investigate the reciprocal interplay between proteins and their surrounding environment. 

In terms of assisting the diagnosis of diseases, the determination of protein mutations could be monitored with THz radiation in addition to the pure detection of unusual molecular concentration levels. For instance, Ebbinghaus et al. have demonstrated that THz spectroscopy can allow for detecting different molecular properties, which are related to solvation dynamics around site-\textcolor{black}{selective} mutated proteins \cite{Ebbinghaus2008}. This is a remarkable result which clearly demonstrates the THz potential for discriminating between well-functioning and defective functional units in the physiological context. \\

\subsection{Heavy biomacromolecules} 
\begin{figure*}
\centering
\includegraphics[width=0.70\textwidth]{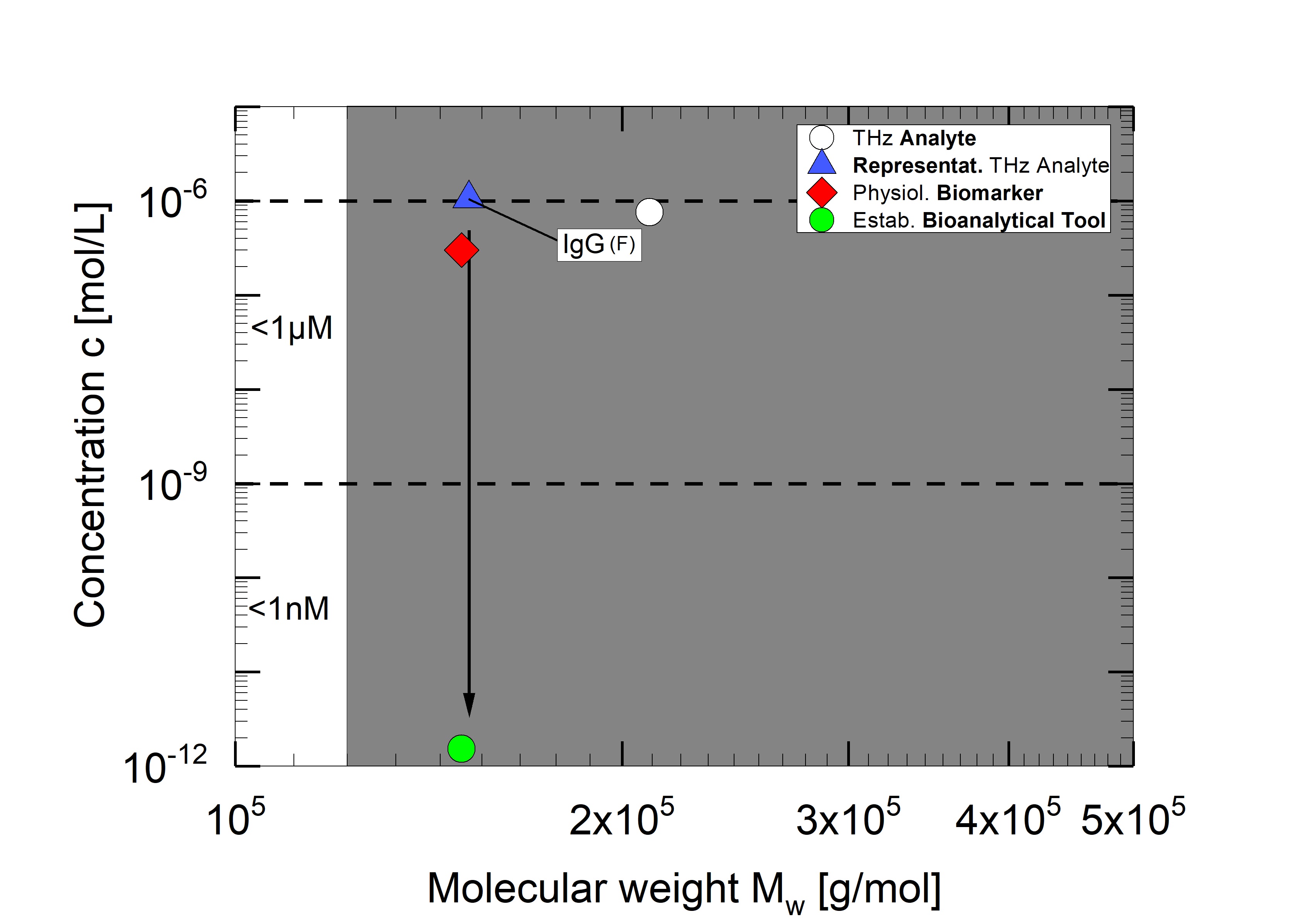}
\caption{\textcolor{black}{Diagram of the reported LDL values of heavy biomacromolecules using traditional THz techniques (white dots), representative analytes (blue triangles), representative physiological concentration levels (red squares) and relevant state-of-the-art bioanalytical techniques (green dots).}}
\label{LDL_Biomacromolecules}
\end{figure*}

One can assume a high complexity in the dielectric response due to the increasing number of excitable molecular modes considering the THz investigation of heavy macromolecules (i.e., large double stranded nucleic acids and conjugated antibodies in this review). Moreover, the interpretation of the dielectric response is expected to become more difficult with increasing molecular complexity. In addition, an increasing complexity of the reciprocal interaction between an analyte with equally increased complexity and its environment (mostly surrounding water molecules) has to be taken into account. This is a big challenge for the characterization of especially large biomacromolecules with THz radiation. On the other hand, considering solely the LDL, both the molecular weight and the complexity are rather advantageous because of an increased amount of charge and oscillating vibrational modes. \textcolor{black}{An overview of the existing investigations on heavy macromolecules is shown in Figure\,\ref{LDL_Biomacromolecules}.}\\

\subsubsection{Subject of THz studies}
The change in dielectric response as a function of an antibodies structure, which is modified by fluorescence and peroxidase conjugates, has been recorded in aqueous environments. Moreover, the optical constants that depend on the modified antibody concentrations have been determined \cite{Sun2011}. In addition, the dielectric response of aqueous embedded double-stranded DNA sequences of significant base-pair lengths producing higher molecular weights M$_w\,>\,10^5\,$g/mol has been investigated \cite{Arora2012, Wu2012}. \\

\subsubsection{(Patho-)physiological relevance and concentration level of IgG antibodies}
Immunoglobulin G (IgG) has been selected to function as representative bioanalyte for the group of heavy biomacromolecules. The IgG antibody production is considered as primary response to bacterial or viral infections. Generally, concentration values that are out of reference ranges can be associated with immunodeficiency diseases \cite{Gressner2013}. IgG antibodies are subdivided into the classes IgG1, IgG2, IgG3, and IgG4 according to the gamma heavy chain \cite{Luttmann2014}. Different (amounts of) subclasses of IgGs are  synthesized depending on the type of antigen, place of incursion, and time of exposition \cite{Gressner2013}. Deficiencies in certain subclasses can indicate infections of the respiratory tract or autoimmune diseases. The determination of the overall IgG concentration does not necessarily provide substantial statements regarding possible deficiencies in subclasses. Therefore, individual differentiated concentration dependence is required. The (patho-)physiological reference ranges strongly depend on age and sex. For example, the density reference range for 12–18 year old children is given as follows: IgG1: $\rho$\,=\,2.8\,-\,8.0\,g/L (c\,$\approx\,18.7\,-\,53.3\,\mu$M), IgG2: $\rho$\,=\,1.15\,-\,5.70\,g/L (c\,$\approx\,7.6\,-\,38\,\mu$M), IgG3: $\rho$\,=\,0.24\,-\,1.25\,g/L (c\,$\approx\,1.6\,-\,8.3\,\mu$M) and IgG4: $\rho$\,=\,0.052\,-\,1.25\,g/L (c\,$\approx\,0.3\,-\,8.3\,\mu$M) \cite{Schauer2003}. \\

\subsubsection{THz spectroscopy versus established bioanalytical measurement system for IgG antibodies}
The concentrations of subclasses of IgG have been determined by ELISA \cite{abcamIgG}. The use of such assays helps in identifying the concentration levels far below the physiological reference range. For example, the Human IgG ELISA Kit (ab195215) enables a detection within a density range between $\rho$\,=\,0.23\,\,-\,15\,ng/mL (c\,$\approx$\,1.53 - 100$\,$pM) with a sensitivity of 0.02\,ng/mL (c\,$\approx$\,0.13\,pM), \textcolor{black}{see Figure\,\ref{LDL_Biomacromolecules}}. Although THz detection is less sensitive compared to modern ELISA Kits, it is already reaching physiological concentrations: Sun et al. show a clearly distinguishable signal for the total IgG densities in the range between $\rho$\,=\,160\,mg/L (c\,$\approx\,1.1\,\mu$M) and $\rho$\,=\,270\,mg/L (c\,$\approx\,1.8\,\mu$M) \cite{Sun2011}. Apart from the sensitivity difference, it remains to be demonstrated if THz spectroscopy is capable of distinguishing IgG at the subclass level and \textcolor{black}{selectively} detecting IgG in a complex physiological matrix. \\

\subsubsection{Perspectives of THz sensing of heavy biomacromolecules}
THz spectroscopy is far from being a serious competitor with regard to sensitivity in comparison to established bioanalytical assays in medical diagnostics. However, the conjugate modifications significantly influence the observed concentration-dependent change of the IgG refractive index constants embedded in a polar liquid. This produces an increased contrast that is favorable for enhancing sensitivity and selectivity. It therefore supports detection by significantly decreasing the lower detection limit. More importantly, it demonstrates the potential for conformational macromolecular analysis with THz systems as the collective vibrational modes are associated with with the tertiary structure of proteins and intermolecular interactions. At the same time, the potential for label-free THz detection remains elusive given the broad nature of the electromagnetic response. However, in aqueous environments the discrimination between unlabeled IgG isoforms in the mM-range could be very significant for clinical diagnostics. Therefore, it remains to be shown if THz spectroscopy is capable of distinguishability at the subclass level for the detection of antibody subclass deficiencies.

Sun et al. show that THz techniques have the fundamental suitability to analyze any biomolecular entity, independent of their molecular weight \cite{Sun2011}. For this, a significant change in the dielectric response is a prerequisite. Such a change can be introduced by water substitution in the case of large macromolecules. In addition, generally a considerable amount of charge is carried and exposed by large macromolecules surfaces where they can interact with the water molecules in the adjacent surrounding, significantly influencing the vicinal water dielectric response. Furthermore, the THz response of a more complex molecule itself contributes to the overall spectral response with increased significance. The trend of superlinear decreasing lower detection limit with increasing molecular weight is explained by adding up all these influences. In aqueous environments, the THz analysis of large and heavy analytes require cumbersome numerical analysis of the spectral response to cope with these combined influencing parameters and the high degree of macromolecular complexity, therefore. \\

\textbf{Summary of reviewed biomolecular analytes which have been investigated in aqueous environments} \\
A wide range of biochemically relevant analytes has been reviewed in the ambit of this survey. The comparison of physiologically relevant concentration levels with lower detection limits that have been obtained with THz sensing approaches partially shows an adequate detection limit in several cases, but also a difference of many orders of magnitude in others. In general, it is important to note that established state-of-the-art bioanalytical techniques are in most cases sensitive enough to meet the requirements in the field of clinical diagnostics. THz technologies have to demonstrate additional features to develop a competitive analytic potential, therefore. \textcolor{black}{In this context, it is relevant that THz-based tools are potentially label-free, less complex, more robust, and non-invasive.} At a matured technological level, competitive THz devices can therefore be advantageous in terms of easier sample handling, simpler opportunities for automation, faster analysis and affordability due to the usage of less consumables. \\

\textbf{What is the THz characterization potential for biomedically relevant investigations under physiological conditions?} \\
The THz characterization potential for biomolecules in the biomedical research field is very high because THz excitations are directly and uniquely associated with delocalized vibrational modes. The excitation of both the molecular analyte and its environment can thereby help in discovering structurally mediated biochemical processes and their sub-ps dynamics. With the THz characterization of hydration shells and the estimation of solvation dynamics, promising results have been obtained. As explained above, the most interesting biological samples for THz examinations are information-carrying molecules, such as oligopeptides and -nucleotides, active agents, proteins and heavy biomacromolecules. To understand the biochemical processes in a much more detailed fashion is a cross-community concern. For gaining more insight into physiological functionalities at the molecular level, the THz community has begun to design and develop suitable analytic tools including advanced numerical deconvolution analysis. Presently most of these THz investigations are carried out under non-physiological conditions but the robustness and performance of THz instrumentation and analytic procedures is becoming mature enough to address such complex environments in the near future.\\

\textbf{What is the key challenge for establishing THz-based analytical techniques for biomolecular detection in aqueous media?} \\ 
Basically, the development of efficient and \textcolor{black}{selective} THz detection principles has been strongly restricted by the ensemble averaging property and the resulting non-specificity of biomolecular THz excitations. This is mostly true for ions, amino acids, nucleotides, active agents, small proteins as well as heavy biomacromolecules. This needs to be addressed, as it is problematic for the development of competitive, fast, reliable and label-free direct detection THz techniques that are suitable as an alternative of the established cross-platform biosensor assays or detection approaches. Many THz analysis methods are based on high analyte contrast and remain to be demonstrated under physiological conditions. Especially, the non-specificity of the THz response strongly limits the potential for \textit{in vivo} real-time applications. Generally, the detection of biomolecular compounds using state-of-the-art bioanalytical tools is a complex procedure that will need to involve also other techniques. Any alternative to shorten preparatory or analytic procedures is strongly desired. It is therefore also plausible to consider and develop THz-radiation-based sensing as a helpful complementary assistive procedure rather than as a complete independent technique replacing established biosensing schemes. In any case, we recommend the THz community to carefully ascertain general statements that imply or even promise fast and complete THz-based sensing solutions in the bioanalytical context without taking full account of the application relevant detection limits and procedural context. \\ 

\textbf{Which one is the next logical step towards THz-assisted sensing in aqueous solutions or even under physiological conditions?}\\
The investigation under physiological conditions is an important step which needs to be addressed. Often, there is a large difference between the sensitivities that are directly accessible with THz sensing techniques and the physiological concentration level as discussed before. Approaches to significantly increase the sensitivity are therefore highly desirable. For instance, as demonstrated for DNA detection, the differential technique is extremely useful to enhance the signal contrast, thereby increasing the sensitivity by several orders of magnitude \cite{Arora2012}. The recently issued dielectric long-range guided mode with absorption could provide another possible construct \cite{sczech2012, Sczech2014}. Possibly, prospective experiments will demonstrate a sensitivity increase due to significantly increased interaction lengths between coupled THz waves and absorbing medium by using this mode. \textcolor{black}{Another promising approach to increase sensitivity is the use of field enhancing elements for example in the form of metamaterials or nanoslot antennas. These have been used for the detection of biomolecules for at least a decade. A good overview of the field of metamaterials is given in the review article by O'Hara et al. \cite{OHara2012}. The nanoslot antenna technology is lucidly summarized in a review article by Adak et al. \cite{Adak2019}. However, these studies and most metamaterial analysis are performed exclusively with dried analytes and thus under non-physiological conditions. The main problem for using metamaterial approaches is the high THz absorption of water which in most designs completely dampens and annihilates the metamaterial´s resonance used for bionsensing. Work recently presented in \cite{Tang2018,Shih2018,Serita2019} addresses this problem by using microfluidic and nanofluidic structures in combination with metamaterials to reduce the influence of water absorption. The result presented in \cite{Serita2019} is two orders of magnitude more sensitive than classical THz techniques and thus demonstrate the great potential of microfluidic metamaterial-based THz sensors. Due to the high interest in this work, we believe that this topic will be the subject of more intensive research in the future and has the potential to significantly increase the sensitivity of THz analyses.} Furthermore, a significant effort to increase \textcolor{black}{selectivity} of the adopted analytic procedures needs to be put in place. For this, a plethora of new approaches is needed, which will most certainly need to be tailored to the specific application field. Differential techniques might have a great potential in this regard, given the unspecific spectra of most biomacromolecules. 

\section{Final remarks on the prospective THz sensing of biochemically relevant analytes in aqueous solution}
The review of THz biomolecular sensing experiments is essentially driven by two major motivations:

Firstly, we consider that after a series of initiatives in the past years, it is time for evaluating not only qualitatively but also quantitatively the potential of THz detection applications in the biomedical context in comparison to established methods. Unfortunately, the direct proclamation of an application potential is stated too early on several occasions, i.e., when a first spectral response of an analyte is detected in an aqueous system, regardless of concentrations orders of magnitude beyond realistic application scenarios. What has been ignored frequently in past reports, is that solutions are required to enhance the \textcolor{black}{selectivity} of the THz response \textcolor{black}{towards actual specificity without} abandoning the appeal of a potential label-free THz sensing technique. This is essential for pursuing the idea of real-time and \textit{in vivo} measurements. Therefore, it is necessary to demonstrate the feasibility of \textcolor{black}{highly selective detection and discrimination} of biomolecules within a pool consisting of structurally similar analytes. This challenge is significant and far from being realized in the near future. We would like to encourage the THz community (and this also includes the authors of the present article) to assess the application situation more precisely, and be careful before THz-based solutions are proclaimed in the biomedical field without carefully analyzing application relevant concentrations, \textcolor{black}{selectivity} in the complex analyte environment, influence of sample variability and alternative techniques.

Secondly, in order to characterize aqueous solutions that are influenced by the interaction with biomolecular solutes, a significant effort is put in the observation of dynamic changes. These approaches are very interesting because they have the potential to better understand mechanisms of signal cascades or communication pathways which are possibly triggered by the biomolecular environment. It is widely accepted that it is necessary to consider the complex biomolecular interplay under real physiological conditions in the era of genomics, proteomics, and glycomics. The logical way of approaching these native conditions has already been addressed by the THz community. The complexity must be increased stepwise by starting from a simple system and the results must be validated by using complementary techniques at all times. Simultaneously, computer-assisted modeling is a reasonable choice to validate experimentally recorded data. In order to reflect the reality, it must be ensured that modeling should be as accurate as possible. The associated risk to oversimplify the situation is high if the modeling assumptions are hypothetical. In reality, the existing models must be precise and the complexity of the experimental situation should not be increased too quickly. 

The characterization of biomolecular (-ly induced) resonances in native aqueous environments is highly desirable after the initial THz investigation of dried and powdered samples. In fact, the complex interpretation of biomolecular measurement results needs to be confirmed by experiments at varied complexity levels and using complementary analytic tools. This is important as we in the THz community are still at the threshold of understanding which basic molecularly induced or solvent related  effects contribute to the measured overall dielectric response in the THz frequency range. By targeting physiological concentrations one can make a relevant step towards analysis under physiological conditions. In this case, a significant increase in sensitivity is required by applying appropriate sensing strategies. The THz investigation by using differential techniques is an option for the THz analysis of biomolecules in aqueous environments at physiological concentrations. 

In addition, we suggest to substantially focus on the \mbox{(patho-)}\-physio\-logical significance of the investigated molecules. Generally, the analysis of biomarkers that are clearly linked to widespread diseases, such as Alzheimer disease, cardiac disease, or cancer is prioritized. In this regard, we also suggest a stronger exchange of information across the THz community borders into relevant neighboring communities (i.e., cell biologists, neuroscientists, oncologists, etc.). This exchange of information is crucial to validate and understand the value and unique potential of THz analysis. In addition, such exchange of information assists in the formulation of a large number of relevant  \mbox{(patho-)}\-physio\-logical research questions. In order to find answers to such questions, the efforts put in may lead us toward the development of more target-oriented THz tools. 

Finally, it is evident from the large increase of publications, that there is a profound interest of the THz community to further explore bioanalytical application and research areas. This critical overview of the present status intends to encourage and guide taking up the relevant and critical steps needed for the development of competitive, substantial and significant THz biosensing tools.

\section*{Acknowledgment}
This work is part of the national priority program SPP 1857 ESSENCE and funded by the Deutsche Forschungsgemeinschaft (DFG, German Research Foundation) under contract numbers HA 3022/8, BO 1573/27 and WI 5209/1.

%
\section*{Conflict of interest}
The authors declare that they have no conflict of interest.

\bibliographystyle{spmpsci}      
\bibliography{manuscript}   

%
%

\end{document}